\documentclass[11pt,a4paper]{article}
\pdfoutput=1
\usepackage{jstyle}

\usepackage{color}
\usepackage{epsfig, palatino}
\usepackage{pstricks,pst-node,pst-tree}
\usepackage{epic}
\usepackage{mathrsfs}
\usepackage{ae} 
\usepackage[T1]{fontenc}
\usepackage[ansinew]{inputenc}
\usepackage{amsmath}
\usepackage{amssymb}
\usepackage{graphicx}
\usepackage{ulem}
\usepackage{color}
\definecolor{darkblue}{cmyk}{0.9,0.9,0,0}
\usepackage{hyperref}
\usepackage{wasysym}
\usepackage{varioref}
\usepackage{makeidx}
\usepackage[english]{babel}
\usepackage{simplewick}
\usepackage{array}
\usepackage{multirow}

\usepackage[font={small}]{caption}

\title{Dissecting supergraviton six-point function with lightcone limits and chiral algebra}

\author[a]{Vasco Gon\c{c}alves,}
\author[b]{Maria Nocchi,}
\author[c]{Xinan Zhou}

\affiliation[a]{Mathematical Institute, University of Oxford, Andrew Wiles Building, Radcliffe Observatory Quar- ter, Woodstock Road, Oxford, OX2 6GG, U.K.}
\affiliation[b]{Centro de Fisica do Porto e Departamento de Fisica e Astronomia, Faculdade de Ciencias da Universidade
do Porto, Rua do Campo Alegre 687, 4169-007, Porto, Portugal}
\affiliation[c]{Kavli Institute for Theoretical Sciences, University of Chinese Academy of Sciences, Beijing 100190, China}

\emailAdd{vasco.dfg@gmail.com}
\emailAdd{nocchi@maths.ox.ac.uk}
\emailAdd{xinan.zhou@ucas.ac.cn}

\abstract{We develop a bootstrap strategy to obtain the six-point function of supergravitons in $AdS_5\times S^5$ from symmetry constraints and consistency conditions. Compared to previous bootstrap algorithms, a novel feature is the use of lightcone OPEs together with the chiral algebra constraint. This makes it possible to isolate different parts of the correlator and fix them separately. Our strategy allows us to gain a refined understanding of the power of different bootstrap constraints, which is also useful for computing more general correlators.}

\begin{document}

\maketitle
\tableofcontents

\newpage
\section{Introduction}
Correlation functions of $\frac{1}{2}$-BPS operators in the four dimensional $\mathcal{N} = 4$ super Yang-Mills (SYM) theory have received a tremendous amount of attention over the past three decades. At large gauge rank $N$ and large 't Hooft coupling $\lambda = g_{YM}^2 N$, these correlators can in principle be computed using Witten diagrams in an expansion around classical type IIB supergravity solution on $AdS_5\times S^5$. In practice, however, this requires a detailed knowledge of complicated effective Lagrangians and it becomes prohibitively hard except for the simplest four-point correlation functions. 

Recent years have witnessed the emergence of a new approach based on symmetries and consistency conditions, which circumvents these computational challenges. 
This is the bootstrap strategy \cite{Rastelli:2016nze, Rastelli:2017udc} which has generated many impressive results.\footnote{See \cite{Bissi:2022mrs} for a review.} For instance,  all infinitely many tree-level  four-point functions of $\frac{1}{2}$-BPS operators with arbitrary Kaluza-Klein (KK) levels have been obtained in all maximally superconformal theories \cite{Rastelli:2016nze, Rastelli:2017udc, Alday:2020lbp, Alday:2020dtb}, as well as in theories with half the amount of maximal superconformal symmetry \cite{Rastelli:2019gtj, Giusto:2020neo, Alday:2021odx}. These results further revealed interesting hidden structures, such as higher dimensional symmetry \cite{Caron-Huot:2018kta,Rastelli:2019gtj,Alday:2021odx}, dimensional reduction \cite{Behan:2021pzk,Alday:2021odx}, and AdS double copy \cite{Zhou:2021gnu}. 

There are many compelling reasons to further extend these studies beyond four points. Holographic correlators are on-shell scattering amplitudes in AdS. As in flat-space, it is important to consider higher particle multiplicities where we expect interesting features to arise.  Another practical motivation is their utility in extracting CFT data at strong coupling. Higher point correlation functions of $\frac{1}{2}$-BPS operators reveal new unprotected data which are not easily accessible through four-point functions alone. This becomes particularly evident when examining the OPE limits. For instance, taking the OPE limit once or twice for a  pair of operators in a six-point function yields five-point or four-point functions containing one or two protected or unprotected double-trace operators respectively. These correlators have attracted attention recently with multi-trace operators being interpreted as bound states in AdS \cite{Ceplak:2021wzz, Bissi:2021hjk, Ma:2022ihn, Aprile:2024lwy,Bissi:2024tqf}. The bootstrap strategy which worked very well for four-point functions also generalizes to higher-point correlators.   For $\mathcal{N}=4$ SYM, the five-point function of the lowest KK mode of supergravitons, i.e., the $20'$ operators, was bootstrapped in \cite{Goncalves:2019znr}. In \cite{Goncalves:2023oyx} the bootstrap approach was further used to obtain an infinite family of five-point functions where two operators have  arbitrary KK levels. Progress has been even more rapid for super Yang-Mills on $AdS_5\times S^3$ \cite{Alday:2021odx,Alday:2022lkk,Alday:2023kfm,Cao:2023cwa,Cao:2024bky,Huang:2024dxr}, partly due to the simpler structure of the theory. In this case, tree-level correlators of supergluons have been determined up to eight points for the lowest KK level  \cite{Cao:2023cwa,Cao:2024bky} and for arbitrary KK modes for five-point functions \cite{Huang:2024dxr}.

These higher-point results were obtained by using methods which follow from two different strategies. The first approach, which led to the results in \cite{Goncalves:2019znr,Alday:2022lkk,Goncalves:2023oyx}, is similar in spirit to the original method of \cite{Rastelli:2016nze, Rastelli:2017udc}. Superconformal symmetry plays a key role in this approach in fixing the ansatz. However, exploiting the superconformal constraints becomes increasingly difficult at higher points. To avoid this difficulty, a second approach was proposed in \cite{Alday:2023kfm} and relies instead on the flat-space limit and factorization properties of AdS amplitudes. This approach is recursive in nature and should ultimately be more efficient. However, to implement it one needs to use lower-point amplitudes as input. These may involve spinning operators other than the scalar supergluons and supergravitons and are not always readily available. Therefore, it is still important to improve the first method in order to maximize our computational power. 

In this paper we present important upgrades of the first method by simplifying the implementation of a key part of superconformal constraints, namely the chiral algebra condition \cite{bllprv13}. The chiral algebra condition is a very nontrivial constraint dictating that correlators in the co-plane configuration with certain twisted R-symmetry polarizations for the operators become meromorphic functions of the complex coordinates on the plane. However, the restriction to the 2d kinematics also makes the condition difficult to use, especially for $n\geq 6$ points. If we use this condition directly in position space, we face a gigantic ansatz which is a complicated function of the cross ratios. In fact, it is even unclear how to write down certain parts of the ansatz explicitly when $n\geq 6$ because we do not know how to evaluate the corresponding Witten diagrams as elementary functions. In Mellin space, the ansatz becomes much simpler. But it is unclear how to use the chiral algebra condition because the Mellin representation requires the operators to be inserted at generic points. Therefore, the use of the chiral algebra constraint has been quite limited in the past works. In this paper we present a new strategy to exploit this condition. The key is to use the lightcone OPE, which together with the chiral algebra condition yield differential equations among different lower-point correlation functions. As we will show, these constraints can be conveniently exploited without leaving Mellin space. Moreover, this new strategy also allows us to isolate different parts of the correlators and examine the consequence of consistency conditions on them separately. This gives us a more refined understanding of how the bootstrap strategy fixes the correlators as compared to earlier implementations where the ansatz is fixed as a whole. To make our discussion more concrete, we will present this improved bootstrap algorithm by considering the tree-level six-point function of $20'$ operators in $\mathcal{N}=4$ SYM. However, it should be obvious that our techniques apply more generally to other correlators and other theories. 

This  paper is organized as follows. In Section \ref{sec:KinMellin} we discuss the superconformal
kinematics of the six-point function, the definition of Mellin representation and flat-space limit of the correlator. In Section \ref{Sec:Mellinfactorization} we review the factorization properties of Mellin amplitudes into lower point functions and explain how the formula for the light-cone OPE can be used to dissect an $n$-point function into products of lower ones. In Section \ref{sec:AnsatzStrategy} we create an ansatz based on the properties explained in the previous sections and explain the main idea of the strategy to fix the correlator. In Section \ref{Sec:snowflake} we illustrate  the implementation of the strategy to fix the correlator by using the chiral algebra twist in a small subsector of the ansatz, called the snowflake piece. In Section \ref{Sec:twosimulpoles} we show that chiral algebra is enough to fix a larger part of the ansatz, called the double pole piece. In Section \ref{Sec:imposeDPtwist} we impose the Drukker-Plefka twist \cite{Drukker:2009sf} and show that this condition, together with the ones implemented in the previous sections are enough to fully determine the six point function of $20'$ operators. We also comment, in this section,   how these twists are satisfied at finite coupling and the implications for lower point functions. In Section \ref{sec:FSL} we check the result obtained in the previous sections against the flat-space limit and obtain a match. In the last Section \ref{sec:DiscussionOutlook} we discuss the result and give an outlook for future directions. In two appendices we collect formulas that were too big for the main text and discuss how the correlator can be expressed in position space.

\section{Kinematics of the six-point function}\label{sec:KinMellin}
\subsection{Implications of superconformal symmetry}
The $20'$ operator, $\mathcal{O}_{20'}^{IJ}$,  is the bottom component of the super multiplet that contains the R-symmetry current $J_{\mu}^{[IJ]}$ and
the stress tensor $\mathcal{T}^{\mu\nu}$ among other operators. It has protected conformal dimension $\Delta =2$ and  transforms in the rank-2 symmetric traceless representation of $SO(6)_R$. Via the AdS/CFT correspondence, this operator is dual to the supergraviton which is a scalar field, while the spinning graviton is dual to $\mathcal{T}^{\mu\nu}$. In this paper, our main focus is the six-point correlation function of $20'$ operators
\begin{align}
\langle \mathcal{O}_{20'}^{I_1J_1}(x_1) \dots \mathcal{O}_{20'}^{I_6J_6}(x_6)  \rangle\;.\label{eq:sixptIndices} 
\end{align}
It is convenient to keep track of the $R$-symmetry indices by contracting them with six dimensional null polarization vectors $y_I$
\begin{align}
\mathcal{O}_{20'}(x,y) \equiv \mathcal{O}_{20'}^{IJ}(x)y_Iy_J\;,  \ \ \ \ \ y^2=0\;. 
\end{align}
This removes all the indices in (\ref{eq:sixptIndices}) and introduces the additional dependence on the R-symmetry coordinates $y_i$
\begin{align}
G_6(x_1,y_1,\dots,x_6,y_6) =\langle \mathcal{O}_{20'} (x_1,y_1) \dots \mathcal{O}_{20'}(x_6,y_6)  \rangle\;.
\end{align}
Clearly, invariance under R-symmetry requires $G_6$ to depend polynomially on $y_{ij}=y_i\cdot y_j$ with degree $2$ for each vector $y_i$. There are $130$ different polynomials which can be generated by different permutations of the following $4$ basic structures\footnote{This should be compared with the cases of four- and five-point functions where there are $2$ independent structures. }
\begin{align}
\nonumber &A^{1}_{i_1i_2i_3i_4i_5 i_6} = y_{i_1i_2}^2y_{i_3i_4}^2y_{i_5i_6}^2\;,  \ \ A^{2}_{i_1i_2i_3i_4i_5 i_6} = y_{i_1i_2}^2y_{i_3i_4}y_{i_4i_5}y_{i_5i_6}y_{i_6i_1}\;, \ \ \\
&A^{3}_{i_1i_2i_3i_4i_5 i_6} = y_{i_1i_2}y_{i_2i_3}y_{i_3i_4}y_{i_4i_5}y_{i_5i_6}y_{i_6i_1}\;, \ \ \ A^{4}_{i_1i_2i_3i_4i_5 i_6} = y_{i_1i_2}y_{i_1i_3}y_{i_2i_3} y_{i_4i_5}y_{i_4i_6}y_{i_5i_6}\;.
\end{align}
Symmetries impose nontrivial constraints on the form of the six-point function. By fully exploiting the bosonic part of superconformal symmetry, i.e., conformal symmetry and R-symmetry, we can write the six-point function in terms of 18 invariant cross ratios. For the spacetime part, we can write down 9 independent conformal cross ratios which we choose to be 
\begin{align}\label{defconfcrossratios}
&u_1=\frac{x_{12}^2x_{35}^2}{x_{13}^2x_{25}^2}\;, \ \ \ \ \ u_{i+1}=u_i\bigg|_{x_j\rightarrow x_{j+1}}\;, \ \ \ U_1=\frac{x_{13}^2x_{46}^2}{x_{14}^2x_{36}^2}\;, \ \ \ U_{i+1}=U_i\bigg|_{x_j\rightarrow x_{j+1}}\;.
\end{align}
Here the points are cyclically identified ($x_7=x_1$) and it is easy to see that the set of conformal cross ratios is $\{u_1,u_2,u_3,u_4,u_5,u_6,U_1,U_2,U_3\}$. Similarly, there are 9 independent R-symmetry cross ratios which can be similarly defined upon replacing $x_{ij}^2\to y_{ij}$. 

The fermionic generators in the superconformal group imposes further constraints which relate the spacetime and R-symmetry dependence. While the full implications are not clear, which presumably can be derived from a superspace analysis, two weaker conditions are known in the literature. The first condition comes from the chiral algebra construction \cite{bllprv13}. When all operators are inserted on a 2d place with complex coordinates $(z_i,\bar{z}_i)$ and R-symmetry polarizations are restricted to the special configuration 
\begin{equation}
    y_{ij}=(z_i-z_j)(v_i-v_j)\;,
\end{equation}
with arbitrary $v_i$, this construction dictates that the correlator becomes independent of the meromorphic coordinates $z_i$
\begin{align}\label{chialg}
G_6(z_i,\bar{z}_i,y_i)\big|_{y_{ij}=(z_i-z_{j})(v_i-v_{j}) } =g(\bar{z}_i,v_i)\;.
\end{align}
The second condition comes from the topological twisting \cite{Drukker:2009sf} which we will refer to as the Drukker-Plefka twist. It was shown in \cite{Drukker:2009sf} that when the R-symmetry polarizations are restricted to 
\begin{equation}
    y_{ij}=x_{ij}^2\;,
\end{equation}
while the locations of the operator insertions are unconstrained, the correlation function becomes topological
\begin{equation}\label{DrukkerPlefkatwist}
G_6(x_i,y_i)\big|_{y_{ij}=x_{ij}^2}={\rm constant}\;.    
\end{equation}
We note here that for four-point functions the Drukker-Plefka twist is implied by the chiral algebra twist by further setting $v_i=\bar{z}_i$ because we can use conformal symmetry to put four points on a plane. This is however no longer the case when we have more than four points and these two conditions are complementary.

\subsection{Mellin representation}\label{Sec:Mellinrepresentation}
An efficient and natural language to describe holographic correlators is the Mellin representation \cite{Mack:2009mi,Penedones:2010ue}. In this formalism, the six-point function of scalars can be represented as 
\begin{align}
 \langle \mathcal{O}_1(x_1,y_1)\dots \mathcal{O}_6(x_6,y_6) \rangle = \int [d\delta]  \prod_{1\leq i <j \leq n} (x_{ij}^2)^{-\delta_{ij}} \Gamma( \delta_{ij})\;  M(\delta_{ij},y_{ij}) \;,    \label{eq:MellinAmpDef}
\end{align}
where $M(\delta_{ij},y_{ij})$ is the Mellin amplitude and the Mellin-Mandelstam variables $\delta_{ij}$ satisfy
\begin{equation}\label{deltaconstr}
   \delta_{ij}=\delta_{ji}\;,\quad\quad \delta_{ii} = -\Delta_i\;,\quad\quad \sum_{j}\delta_{ij} = 0\;.
\end{equation}
The advantage of this formalism is that the analytic structure of the Mellin amplitude is very simple and resembles that of the tree-level scattering amplitudes in flat-space. Moreover, the Mellin amplitude enjoys factorization properties which allow us to express its residues at poles in terms of lower-point Mellin amplitudes \cite{Goncalves:2014rfa}. Since many intermediate lower-point correlators have already been obtained, the factorization of Mellin amplitudes will play an important role in our strategy of computing the six-point function. We will discuss Mellin factorization in more detail in Section \ref{Sec:Mellinfactorization}. 

Another useful property of the Mellin representation is that we can recover the flat-space amplitude from the high energy limit of the Mellin amplitude \cite{Penedones:2010ue}. If we rescale the Mellin-Mandelstam variables by $\delta_{ij}\to\beta \delta_{ij}$, then in the $\beta\to\infty$ limit the leading term of the supergraviton Mellin amplitude gives the flat-space amplitude 
\begin{equation}\label{flatspacerelation}
    \lim_{\beta\to\infty} M(\beta\delta_{ij},y_{ij}) \propto A_{\rm flat}(\delta_{ij},y_{ij})\;.
\end{equation}
On the RHS, we have the tree-level amplitude of gravitons and $\delta_{ij}$ should be identified with the flat-space Mandelstam variables $\delta_{ij}=p_i\cdot p_j$. However, the flat-space amplitude obtained in this way is not in the most generic kinematic configuration. This arises essentially because of the factorized structure of $AdS_5\times S^5$ and supergravitons are essentially 10d gravitons with polarizations pointing along the internal $S^5$. In flat-space, each graviton is associated with a polarization tensor $\epsilon_{\mu\nu}$. Obtained from AdS, these flat-space polarization tensors are related to the R-symmetry polarization tensors as \cite{Chester:2018aca,Alday:2021odx}
\begin{equation}
    \epsilon_{\mu\nu}=y_\mu y_\nu\;,
\end{equation}
where we lift the six dimensional $y_I$ into ten dimensional $y_\mu$ by adding zeros.

\section{Mellin factorization and OPE}\label{Sec:Mellinfactorization}

\subsection{Factorization of Mellin amplitudes}
Mellin amplitudes are analytic functions with poles in $\delta_{ij}$ associated with exchanged single-trace operators in the OPE. More concretely, let us separate the external operators into two subsets $L$ and $R$ and consider the exchange of an operator with dimension $\Delta$ and spin $J$ between these two groups of operators. The Mellin amplitude has poles at
\begin{align}
M(\delta_{ij},y_{ij}) \approx \sum_m \frac{Q_m(\delta_{ij},y_{ij})}{\delta_{LR} -(\Delta-J+2m)}\;,  \ \ \ \delta_{LR} =\sum_{i\in L}\sum_{j \in  R}\delta_{ij}\;,\label{eq:factorizationMellin}
\end{align} 
where the poles with $m>0$ correspond to descendant operators. For generic operator dimensions, the sum over $m$ runs from $0$ to $\infty$. However, for the particular theory under consideration, we will see in the appendix that the infinite series of poles truncates to a finite set (the mechanism is similar to one already observed for five-point functions \cite{Goncalves:2023oyx}). An interesting property of Mellin amplitudes is that the residues at these poles are related to lower-point amplitudes \cite{Goncalves:2014rfa}, which resembles amplitude factorization in flat-space. Here we review this factorization property. For the simplicity of presentation, we will suppress the R-symmetry dependence which obeys its own ``factorization'' and can be multiplied back in the end. We will discuss the consequence of operator exchanges on R-symmetry structures in Section \ref{sec:rsymmetry}.

The precise factorization formula for the residues depends crucially on the spin of the exchanged operator. Let us first introduce the extension of the scalar Mellin amplitude (\ref{eq:MellinAmpDef}) where one of the operators becomes spinning. The correlator of $n$ scalar operators and a spinning operator with dimension $\Delta$ and spin $J$ can be represented as \cite{Goncalves:2014rfa}
\begin{equation}\label{defspinningMellin}
\begin{split}
\langle \mathcal{O}(x_0,z) {\mathcal{O}}_1(x_1) \dots {\mathcal{O}}_n(x_n) \rangle  ={}& \sum_{a_1,\dots a_J=1}^n \prod_{\ell=1}^{J} (z\cdot x_{0a_\ell})  \int [d\gamma] \;M^{\{a\} }(\gamma_{ij},\gamma_i)\\
{}&\quad\quad\quad\quad\times\prod_{1\leq i<j\leq n} \frac{\Gamma(\gamma_{ij} )}{(x_{ij}^2)^{\gamma_{ij} }} \prod_{i=1}^n \frac{\Gamma(\gamma_{i} +\{a\}_i )}{ (x_{i0}^2)^{\gamma_{i} +\{a\}_i } }\;,   
\end{split}
\end{equation}
where $z$ is the polarization vector of the spinning operator. We have collectively denoted $a_1,a_2,\ldots,a_J$ by $\{a\}$ and introduced $\{a\}_i$ to count the number of times the index $i$ appears in $\{a\}$
\begin{equation}
   \{a\}_i = \delta_{a_1}^i+\delta_{a_2}^i+\ldots+\delta_{a_J}^i\;. 
\end{equation}
Similar to (\ref{deltaconstr}), conformal invariance requires the Mellin-Mandelstam variables to satisfy
\begin{align}
\gamma_{ij}=\gamma_{ji}\;,\quad \gamma_{ii} = -\Delta_i\;,\quad \gamma_i = - \sum_{j=1}^n \gamma_{ij}\;,\quad \sum_{i,j=1}^n\gamma_{ij}=J-\Delta
\;.
\end{align}
The spinning Mellin amplitudes $M^{\{a\}}$ are not all independent but  satisfy the transversality condition \cite{Goncalves:2014rfa}
\begin{align}
\sum_{a_1=1}^n (\gamma_{a_1} + \delta_{a_1}^{a_2} + \delta_{a_1}^{a_3} + \cdots + \delta_{a_1}^{a_J})M^{a_1 a_2 \cdots a_J} = 0\;.
\end{align}
The residue of the amplitude (\ref{eq:factorizationMellin}), for $m=0$, can be expressed as \cite{Goncalves:2014rfa}
\begin{align}
Q_0(\delta_{ij}) =k_{\Delta,J} \sum_{{a} \in L} \sum_{i\in R} M_L^{\{a\}} M_R^{\{i\}} \prod_{\ell=1}^J (\delta_{a_\ell i_\ell} + \delta_{a_\ell}^{i_{\ell+1}}\delta_{i_\ell}^{i_{\ell+1}} + \cdots + \delta_{a_\ell}^{i_J}\delta_{i_\ell}^{i_{\ell+J}} )\;,
\end{align}
where $k_{\Delta,J}$ is a normalization constant, $M_L^{\{a\}}$, $M_R^{\{i\}} $ are lower-point Mellin amplitudes and $\{a\}=a_1\dots a_J,\ \{i\}=i_1\dots i_J$. For $m>0$, so far there is no general formula valid for any spin.  However, residue formulas for spins up to two have been obtained in \cite{Goncalves:2014rfa} and are sufficient for the purpose of this work.  Here we will not present the detailed expressions which are collected in the appendix. It is sufficient to point out that the residues with $m>0$ are determined by conformal symmetry. A convenient way to implement conformal symmetry is to use the conformal Casimir which was first used in \cite{Dolan:2003hv} to compute four-point conformal blocks. The Casimir used here is a multi-particle operator constructed from the sum of conformal generators acting on the operators in the group $L$ (we will give more details in Section \ref{sec:rsymmetry} for the similar case of R-symmetry). By simple arguments based on conformal invariance (see, e.g., \cite{zhou:2018sfz}), the Casimir operator is mapped to the Laplacian operator in AdS which collapses the propagator of the exchanged field to a delta function. In position space, the Casimir operator is a differential operator acting on the coordinates. Acting on the definition (\ref{eq:MellinAmpDef}) and shifiting $\delta_{ij}$, we find it gets translated into a difference operator. The action of the difference operator removes the poles in $\delta_{LR}$ in (\ref{eq:factorizationMellin}) and relates the residues for different $m$ via recursion relations. In fact, such recursion relations not only determines the residues for $m>0$ in terms of those for $m=0$ but also places constraints on the residues with $m=0$.     

Let us now specialize to the case of the six-point supergraviton amplitude. There are two nontrivial factorizations where the six-point amplitude splits into the product of three-point and five-point amplitudes or the product of two four-point amplitudes\footnote{This should be contrasted with the five-point amplitude case where there is only one nontrivial way to factorize it into three-point and four-point amplitudes.}
\begin{align}
&M (\delta_{ij}) \approx \frac{Q_m^{(4-4)}(\delta_{ij})}{ (\delta _{l_1r_1}+\delta _{l_1r_2}+\delta _{l_1r_3}+\delta _{l_2r_1}+\delta _{l_2r_2}+\delta _{l_2r_3}+\delta _{l_3r_1}+\delta _{l_3r_2}+\delta _{l_3r_3}) -(\tau+2m)}\;,\label{eq:combtype}\\
&M (\delta_{ij}) \approx  \frac{Q_m^{(3-5)}(\delta_{ij})}{ (\delta _{l_1r_1}+\delta _{l_1r_2}+\delta _{l_1r_3}+\delta _{l_1r_4}+\delta _{l_2r_1}+\delta _{l_2r_2}+\delta _{l_2r_3}+\delta _{l_2r_4}) -(\tau+2m)}\;.\label{eq:snowflaketype}
\end{align} 
Here $l_i$ and $r_i$ belong to different groups of external operators and we have added superscripts to the residues to indicate their factorizations into lower-point amplitudes. In the factorization channel, only three operators can appear in our setup. These are the $20'$ operator itself, the R-symmetry current and the stress tensor operator. Note that all these operators have the same conformal twist $\tau=2$, which provides simplification to the pole structure of the Mellin amplitude. 

To determine the residues in  (\ref{eq:combtype}), one would need the following four-point functions 
\begin{align}
& \langle {\mathcal{O}}_{20'}(x_1,y_1) {\mathcal{O}}_{20'}(x_2,y_2) {\mathcal{O}}_{20'}(x_3,y_3) {\mathcal{O}}_{20'}(x_0,y_0) \rangle\;, \\ & \langle {\mathcal{O}}_{20'}(x_1,y_1) {\mathcal{O}}_{20'}(x_2,y_2) {\mathcal{O}}_{20'}(x_3,y_3) \mathcal{J}(x_0,z_0) \rangle\;, \nonumber\\
&\langle {\mathcal{O}}_{20'}(x_1,y_1) {\mathcal{O}}_{20'}(x_2,y_2) {\mathcal{O}}_{20'}(x_3,y_3) \mathcal{T}(x_0,z_0) \rangle\;.
\end{align}
 Fortunately, these have already been computed in \cite{Belitsky:2014zha,Goncalves:2019znr} and we collect their formulas in the appendix. By contrast, to compute the residues in (\ref{eq:snowflaketype}) the following five-point functions are needed
\begin{align}
&\langle {\mathcal{O}}_{20'}(x_1,y_1) {\mathcal{O}}_{20'}(x_2,y_2){\mathcal{O}}_{20'}(x_3,y_3){\mathcal{O}}_{20'}(x_4,y_4) {\mathcal{O}}_{20'}(x_0,y_0) \rangle\;, \\ & \langle {\mathcal{O}}_{20'}(x_1,y_1) {\mathcal{O}}_{20'}(x_2,y_2){\mathcal{O}}_{20'}(x_3,y_3){\mathcal{O}}_{20'}(x_4,y_4) \mathcal{J}(x_0,z_0) \rangle\;, \nonumber\\
& \langle {\mathcal{O}}_{20'}(x_1,y_1) {\mathcal{O}}_{20'}(x_2,y_2){\mathcal{O}}_{20'}(x_3,y_3){\mathcal{O}}_{20'}(x_4,y_4) \mathcal{T}(x_0,z_0) \rangle\;.
\end{align}
While the supergraviton five-point function has been computed in \cite{Goncalves:2019znr}, the other two spinning correlators are not known. Unlike the four-point function case where the spinning correlators are related to the scalar correlator by superconformal symmetry, this is not true for five-point functions \cite{Heslop:2022xgp}.\footnote{However, these five-point functions can be obtained as a byproduct of our analysis since the six-point function will be completely fixed.} 

In this subsection, we have only applied factorization to a single channel in the supergraviton Mellin amplitude. Conceptually, there is no difficulty to further consider factorization on two or more compatible channels. However, the problem is that this would generically involve lower-point functions with multiple spinning operators. It is not clear how to further extend the Mellin representation (\ref{defspinningMellin}) to handle the generic case and this makes the direct generalization of the factorization analysis technically infeasible. We will not present a general solution in this paper. However, in the next two subsections we will provide a way to implement multi-factorization within the scope of our problem.

 \subsection{Multiple factorization: Two spinning lines}

To handle multiple factorizations, we start from position space. The main ingredient we need is the following  formula which implements the lightcone OPE between two scalar identical operators \cite{Bercini:2020msp}
\begin{align}
&\mathcal{O}(x_1)\,\mathcal{O}(x_2) \approx  \sum_{J}C_{12J} \int_{0}^{1}  [dt]\,\frac{\mathcal{O}_{J}(x_1+tx_{21},x_{12}) }{(x_{12}^2)^{\frac{2\Delta_{\mathcal{O}}-\tau}{2}}} \label{eq:LightconeOPEFormula}
+\dots\;. 
\end{align}
Here $\tau=\Delta-J$ is the twist of the exchanged operator and $C_{12J}$ is the OPE coefficient. In the lightcone limit $x_{12}^2\to 0$ and the polarization vector of the spinning operator is taken to be the null vector $x_{12}$. The lightcone OPE involves an integral and the integration measure is given by $[dt] =dt\, (t(1-t))^{\frac{\Delta+J-2}{2}}$.\footnote{The generalization to different external operators is simple. One just needs to change $2\Delta_{\mathcal{O}}\rightarrow \Delta_1+\Delta_2$ and $[dt] \rightarrow dt (t(t-1))^{\frac{\Delta+J-2}{2}} t^{\frac{\Delta_{1}-\Delta_{2}}{2}}(1-t)^{\frac{\Delta_2-\Delta_1}{2}}$.} As we commented in the previous subsection, the conformal Casimir is very useful in the context of factorization and acts diagonally on the factorization channel. This property has automatically been taken into account by (\ref{eq:LightconeOPEFormula}).

To go beyond single factorization, we can apply the lightcone OPE multiple times. Let us first apply it twice to reduce the six-point function into four-point functions. We get
\begin{align}
\langle \mathcal{O} (x_1)\dots &\mathcal{O} (x_6) \rangle\big|_{ x_{12}^2 \to 0 \atop x_{34}^2\rightarrow 0}  =  \sum_{J_1,J_2} \frac{C_{12J_1}C_{34J_2} }{(x_{12}^2x_{34}^2)^{\Delta_{\mathcal{O} }}}\label{eq:twolightconespos} \\
&\times\int [dt_1][dt_2] (x_{12}^2x_{34}^2)^{\frac{\tau}{2} } \langle  \mathcal{O}_{J_1}(x_2+t_1x_{12},x_{12})\mathcal{O}_{J_2}(x_4+t_2x_{34},x_{34}) \mathcal{O} (x_5)\mathcal{O} (x_6) \rangle\;, \nonumber
\end{align}
where we have assumed that the leading contribution comes from a family of operators with twist $\tau$.

It is well known that conformal symmetry allows us to write four-point correlators as functions of two conformal cross ratios. Importantly, the cross ratios in the four-point functions in (\ref{eq:twolightconespos}) can be expressed in terms of the six-point cross ratios introduced in (\ref{defconfcrossratios}). As a result, we get the following integral
\begin{align}
&\int [dt_1][dt_2]\frac{u_1^{\tau /2} u_3^{\tau /2} (U_1-u_2)^{\ell } U_2^{k_1+2 \ell -J_1} (t_2(U_1+U_2-u_2 U_2-1)-U_2+1)^{k_1}}{ u_5^{-k_1-k_2-2-\tau -\ell } }\label{eq:twopolesblockposition} \\
& (t_1 u_4 (U_1-u_2 U_2)+t_1 (U_2-1) U_3+U_2 (u_2 u_4-U_3))^{k_2} (U_2-u_6)^{J_1-k_1-\ell } (U_3-u_4)^{J_2-k_2-\ell } \nonumber\\
&\frac{(t_2U_3+u_4-t_2u_4)^{J_1-J_2-k_1} U_1^{-k_1-k_2-\frac{\tau }{2}-2 \ell }}{ X_1^{J_1+J_2+\tau } (x_{56}^2)^{\Delta_{\mathcal{O}}}} f_{k_1k_2\ell }^{J_1J_2}(X_1,X_2)\;, \nonumber
\end{align}
where $k_i$ and $\ell$ are summed over, as explained below, and 
\begin{equation}
    \begin{split}
        X_1={}&\frac{u_5 (t_1 t_2(U_1+U_2-u_2 U_2-1)+t_1+(1-t_1-t_2+t_2u_2) U_2)}{U_1}\;,\\
       X_2={}& \frac{(t_1 (u_6-U_2)+U_2) ((1-t_2) u_4+t_2U_3)}{U_2}\;.
    \end{split}
\end{equation}
To be precise, we have used 
\begin{align}
&\langle \mathcal{O}_{J_1}(x_2,z_2)  \mathcal{O}_{J_2}(x_4,z_4) \mathcal{O}(x_5)\mathcal{O}(x_6) \rangle=\label{eq:fourpointegrals1}\\
&=\sum_{k_i,\ell}\frac{H_{24}^{\ell } V_{2,45}^{k_1} V_{4,25}^{k_2} V_{2,56}^{J_1-k_1-\ell } V_{4,56}^{J_2-k_2-\ell } }{\left(x_{56}^2\right)^{\Delta_{\mathcal{O} }  } \left(x_{24}^2\right)^{\frac{2\tau+J_1+J_2 }{2}  }  } \left(\frac{x_{25}^2}{x_{45}^2}\right)^{\frac{J_2-J_1}{2} } f_{k_1k_2\ell }^{J_1J_2}\left(\frac{x_{24}^2 x_{56}^2}{x_{25}^2 x_{46}^2},\frac{x_{26}^2 x_{45}^2}{x_{25}^2 x_{46}^2}\right)\;,\nonumber
\end{align}
where $J_1$, $J_2$ are the two exchanged spins and we sum over all possible tensor structures of the spinning four-point function \cite{Costa:2011mg} (see Appendix \ref{eq:estruturas3pf} for the definitions of $H$ and $V$ structures).
The powers of $(x_{12}^2)^{\frac{2\Delta_{\mathcal{O}}- \tau }{2}}$ and $(x_{34}^2)^{\frac{2\Delta_{\mathcal{O}}- \tau }{2}}$ (or equivalently the powers of $u_1$ and $u_3$) can be obtained, from the Mellin amplitude point of view, by taking the residues at
\begin{align}
\delta_{12} = \frac{2\Delta_{\mathcal{O} }-\tau }{2}\;,  \ \ \ \delta_{34} = \frac{2\Delta_{\mathcal{O} }-\tau }{2}\;.\label{eq:residues}
\end{align}
The form of (\ref{eq:fourpointegrals1}) motivates us to write the functions $f_{k_1k_2\ell }^{J_1J_2}$ in terms of Mellin type integrals
\begin{align}
f_{k_1k_2\ell }^{J_1J_2}(u,v) = \int [dt] [ds] M_{k_1k_2\ell }^{J_1J_2}(s,t) u^{s}v^{t}\;,\label{eq:fourptMellinSpin1}
\end{align}
which makes (\ref{eq:fourpointegrals1}) reminiscent of the Mellin representation. This connection can be made even more manifest by using  identities of the following type
\begin{align}
   ((1-t_i) u_a+t_i U_b)^{\beta}= \int [d\delta] \frac{  \Gamma (\delta) \Gamma (-\beta-\delta) (t_i U_b)^{\beta+\delta}}{u_a^{\delta}(1-t_i)^{\delta}\Gamma (-\beta)}\;,\label{eq:Mellinintid}
\end{align}
where the integral over $\delta$ is along the imaginary axis and is nothing more than the Mellin transform with respect to the variable $u_a$ (the Mellin transform with respect to $U_b$ can be obtained similarly). In analyzing the $X_2$ term, we apply this Mellin transform to each of the two factors with respect to their $u_a$ variables. For $X_1$, we need to use this identity three times. First, we rewrite the expression in $X_1$ using a Mellin transform with respect to $U_2$
\begin{align}
    &(t_1 t_2(U_1+U_2-u_2 U_2-1)+t_1+(1-t_1-t_2+t_2u_2) U_2)^{\beta} \nonumber\\
    &= ((1-t_1)(1-t_2+t_2u_2)U_2+t_1(1-t_2+t_2U_1))^\beta \nonumber\\
    &= \int [d\delta] \frac{\Gamma (\delta )\Gamma (-\beta -\delta ) (1-t_1)^{-\delta }  t_1^{\beta +\delta } }{U_2^\delta \Gamma (-\beta )}(1-t_2+t_2 u_2)^{-\delta } (1-t_2+t_2 U_1)^{\beta +\delta }\;.
\end{align}
 Then, it is clear that we can do two more Mellin transforms for the last two factors. The use of (\ref{eq:Mellinintid}) in the previous analysis introduces five Mellin integrals and  it trivializes integrals over $t_1$ and $t_2$. The result can be written in terms of only powers of the cross ratios. Moreover, note that the five Mellin type integrals from using the identity, plus two more from (\ref{eq:fourptMellinSpin1}) and another two already used in taking residues at  (\ref{eq:residues}), add up to $9$, which is precisely the number of independent Mellin integrals for a six-point function. While the outlined procedures for writing (\ref{eq:twolightconespos}) is straightforward, the exact details are too lengthy to write down here explicitly. For this reason, we only quote the final result
\begin{align}
    (\ref{eq:twolightconespos}) \rightarrow M(\delta_{ij})\big|_{\delta_{12},\delta_{34}\to 1} \approx \sum_{J_1J_2,k_1k_2,\ell}\frac{C_{12J_1}C_{34J_2}Q_{k_1k_2\ell }^{J_1J_2}(\delta _{16},\delta _{23},\delta _{24},\delta _{45},\delta _{46},s,t) }{(\delta_{12}-1)(\delta_{34}-1)}\;,
\end{align}
where $s=2-\delta_{56}, \ t= -(\delta_{16}+\delta_{26})$. The function $Q_{k_1k_2\ell }^{J_1J_2}$ is polynomial in the first five arguments while the last two enter through the four point Mellin amplitude $M_{k_1k_2\ell }^{J_1J_2}(s,t)$. For example, $Q_{001}^{22}$ is given by
\begin{equation}
\begin{split}
  Q_{001}^{22} = {}& -\frac{225 (\delta_{24} (s-1-\bar{\delta}_{456})-\delta_{23} (\bar{\delta}_{456}-1)) }{4 \Gamma (2-s) \Gamma (4-s) \Gamma (1-t)^2 \Gamma (s+t+1)^2} \\ 
  {}&\times (s \delta_{16}+t (\bar{\delta}_{234}+s-1)) (s \delta_{45}+t \bar{\delta}_{456}) M_{001}^{22}(s+2,t-1)\;,
  \end{split}
\end{equation}
where $\bar{\delta}_{ijk} = \delta_{ij}+\delta_{ik}$. We provide other cases in an auxilliary file. 

\subsection{Multiple factorization: Three spinning lines}\label{Sec:fac3spinning}
The analysis of the previous subsection can be straightforwardly extended to the case where we have three simultaneous lightcone OPEs. The generalization of (\ref{eq:twolightconespos}) to this case is given by
\begin{align}
&\langle \mathcal{O}(x_1)\dots \mathcal{O}(x_6)\rangle=\frac{1}{(x_{12}^2x_{34}^2x_{56}^2)^{\Delta_{\mathcal{O} }}}\sum_{k_i}q_{k_1k_2k_3}G_{\mathcal{O}_{k_1}\mathcal{O}_{k_2}\mathcal{O}_{k_3}}(u_i,U_i)\label{eq:deflightconeblocks}\\
&\to\sum_{J_i}  \left( \prod_{i=1}^3  C_{\mathcal{O}\mathcal{O} J_i}\int [dt_i]\right) 
\frac{\langle \mathcal{O}_{J_1}(x_1+t_1x_{21},x_{12})\mathcal{O}_{J_2}(x_3+t_2x_{43},x_{34})\mathcal{O}_{J_3}(x_5+t_3x_{65},x_{56})  \rangle}{(x_{12}^2x_{34}^2x_{56}^2)^{\frac{2\Delta_{\mathcal{O}}-\tau}{2}} }\,.\nonumber
\end{align}
Here, the first line is a schematic conformal block decomposition of the correlator under a triple OPE. The expression in the second line is valid at the leading order of the lightcone limit $x_{12}^2, x_{34}^2, x_{56}^2\to 0$.
Since in the supergravity limit only twist 2 operators are exchanged and they all have different spins, we used their spins to label different contributions to the correlator. Similar to (\ref{eq:fourpointegrals1}), the correlator can be decomposed into different tensor structures
\begin{equation}
    \langle \mathcal{O}_{J_1}\mathcal{O}_{J_2}\mathcal{O}_{J_3}\rangle = \sum_{\ell_1,\ell_2\ell_3}\frac{C_{J_1J_2J_3}^{\ell_1\ell_2\ell_3} V_{1,23}^{J_1-\ell_2-\ell_3}V_{2,13}^{J_2-\ell_1-\ell_3}V_{3,12}^{J_3-\ell_1-\ell_2}H_{12}^{\ell_{3}} H_{13}^{\ell_{2}}H_{23}^{\ell_{1}}}{(x_{12}^2)^{\frac{\tau+2(J_1+J_2-J_3)}{2}}(x_{13}^2)^{\frac{\tau+2(J_1+J_3-J_2)}{2}}(x_{23}^2)^{\frac{\tau+2(J_2+J_3-J_1)}{2}}}\;,\label{eq:threepointfunctionstructure}
\end{equation}
where we assumed that the operators have the same twist $\tau$.

The second line in (\ref{eq:deflightconeblocks}) together with the structure of the three-point function (\ref{eq:threepointfunctionstructure}) gives a useful representation of the conformal block $G_{J_1J_2J_3}^{\ell_1\ell_2\ell_3}(u_i,U_i)$ in the lightcone limit. The dependence on the cross ratios $u_1,u_3$ and $u_5$ trivializes
\begin{align}
G_{J_1J_2J_3}^{\ell_1\ell_2\ell_3}(u_i,U_i)\rightarrow (u_{1}u_3u_5)^{\frac{\tau}{2}}g_{J_1J_2J_3}^{\ell_1\ell_2\ell_3}(u_{2i},U_i)\;,\label{eq:lightconeblocklimit}
\end{align}
and $g_{J_1J_2J_3}^{\ell_1\ell_2\ell_3}$ is a function of six cross ratios whose expression we record in the appendix. Our goal now is to repeat the same  strategy that was implemented in the last subsection and obtain a map between the triple integral representation of $G_{J_1J_2J_3}^{\ell_1\ell_2\ell_3}(u_i,U_i)$ and the Mellin amplitude. As one can see from (\ref{eq:conformalblocklightcone}) in the appendix, the function $g_{J_1J_2J_3}^{\ell_1\ell_2\ell_3}$ has a simple integral expression. In the following we will use the  Mellin transform  formula (\ref{eq:Mellinintid}) to simplify the dependence on the cross ratios $u_a$ and $U_a$. There are three nontrivial terms involved in this Mellin transformation (the $B_i$ in (\ref{eq:conformalblocklightcone}) ). One of the terms has the form
\begin{align}
    ((1-t_1) t_2 u_2U_2+U_1 (t_1 t_2 U_2-t_1 U_2+t_1-t_2 U_2+U_2))^{\ell_1+\ell_2-J_1-J_2-\frac{\tau }{2}}\;,\label{eq:exampleMellinsnowflake6pt}
\end{align}
and the other two are obtained by permutation. The next step is to do a Mellin transformation with respect to $u_2$ using the formula (\ref{eq:Mellinintid})
\begin{align}
(\ref{eq:exampleMellinsnowflake6pt}) \to \int [d\delta] \frac{\Gamma (\delta )  \Gamma \left(J_1+J_2-\ell _1-\ell _2-\delta +\frac{\tau }{2}\right) \left(\left(t_1-1\right) \left(t_2-1\right) U_2+t_1\right){}^{\delta -J_1-J_2-\frac{\tau }{2}+\ell _1+\ell _2} }{ U_1^{ J_1+J_2+\frac{\tau }{2}-\ell _1-\ell _2- \delta} (u_2U_2t_2(1-t_2))^{\delta }  \Gamma \left(\frac{\tau }{2}+J_1+J_2-\ell _1-\ell _2\right)}\nonumber
\end{align}
and then another one with respect to $U_2$ on the last term in the numerator of the previous expression
\begin{align}
&\left(\left(t_1-1\right) \left(t_2-1\right) U_2+t_1\right){}^{\delta -J_1-J_2-\frac{\tau }{2}+\ell _1+\ell _2} \\
&=\int [d\bar{\delta}] \frac{\Gamma (\bar{\delta})  t_1^{\delta +\bar{\delta}-J_1-J_2-\frac{\tau }{2}+\ell _1+\ell _2} \Gamma \left(\frac{\tau }{2}+J_1+J_2-\delta -\bar{\delta}-\ell _1-\ell _2\right)}{  \left(U_2\left(1-t_1\right) \left(1-t_2\right)\right)^{\bar{\delta}} \Gamma \left(\frac{\tau }{2}+J_1+J_2-\delta -\ell _1-\ell _2\right)}\nonumber
\end{align}
which trivializes the dependence on the cross ratios and allows for the integration over the  $t_i$ variables.  Since there are two other factors like (\ref{eq:exampleMellinsnowflake6pt}), we get six Mellin type integrals in total. Three other integrals can be introduced by picking poles that reproduce the powers of $u_1,u_3$ and $u_5$ in  (\ref{eq:lightconeblocklimit}). These steps lead to the following map
\begin{align}\label{mapGM3pt}
    G_{J_1J_2J_3}^{\ell_1\ell_2\ell_3}(u_i,U_i)\bigg|_{\textrm{light-cone}}\rightarrow \frac{M_{J_1J_2J_3}^{\ell_1\ell_2\ell_3}(\delta)}{(\delta_{12}-1)(\delta_{34}-1)(\delta_{56}-1)}\;,
\end{align}
where the residues $M_{J_1J_2J_3}^{\ell_1\ell_2\ell_3}(\delta)$ can be viewed as the generalization of Mack polynomials to six-point functions (analogous formulas for five-point Mellin amplitude were derived in Appendix A of \cite{Antunes:2021kmm}). For this work, we will only need $M_{J_1J_2J_3}^{\ell_1\ell_2\ell_3}$ for spins up to $2$. These expressions are quite cumbersome and we relegate them to the appendix. 
Before concluding this subsection, let us comment that the extension to correlators with more than six operators is completely analogous. If we start with a scalar $2n$-point function, we can perform at most $n$ lightcone OPEs and relate the residue of $n$ poles of the Mellin amplitude to a spinning $n$-point function. If we instead start with a scalar $(2n-1)$-point function, we can perform at most $n-1$ lightcone OPEs to relate the residue of $n-1$ poles of the Mellin amplitude to an $n$-point function where $n-1$ operators are spinning. Although it is not clear what the best definition is for general spinning Mellin amplitudes, for the purpose of exploiting the information the lower-point correlators it is sufficient to decompose them into tensor structures and Mellin transform the scalar functions. The map between the multi-lightcone OPE limit in position space and the residues in Mellin space can be obtained in a similar way. We hope that this strategy will be used for more than six points in the future. 

\subsection{R-symmetry polynomials} \label{sec:rsymmetry}

In the previous subsections we studied how the Mellin amplitude factorizes into lower-point amplitudes at its poles. These poles correspond to the exchanged single-trace operators which transform in specific R-symmetry representations. In this subsection, we study how the information of exchanged R-symmetry representations is encoded in the R-symmetry depdendence of the correlator.

\begin{figure}
 	\centering
 	\includegraphics[width=0.8\linewidth]{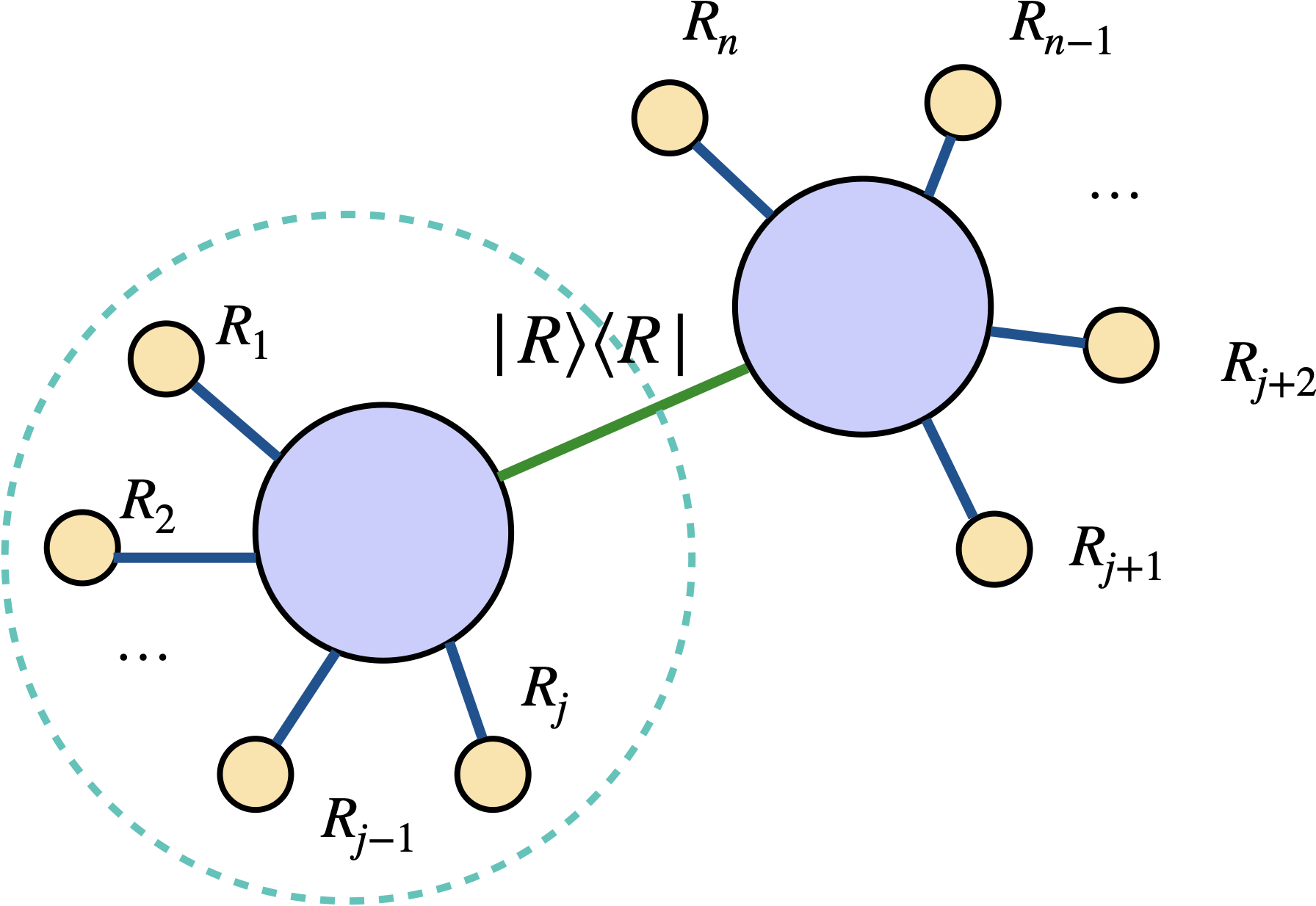}
 	\caption{Exchange of an R-symmetry representation.}
 	\label{fig:Rsymm}
 \end{figure}

To study them, we can insert a projector which allows only a certain irreducible representation to propagate in the exchange channel (see Figure \ref{fig:Rsymm}). It is clear that the part encircled by the green dashed line circle is invariant under R-symmetry transformations. In other words, the sum of actions of R-symmetry generators on the first $j$ external operators can be traded for the action on the internal projector 
\begin{equation}
    L^{(1)}_{IJ}+L^{(2)}_{IJ}+\ldots+L^{(j)}_{IJ}=L^{\rm int}_{IJ}\;.
\end{equation}
Using this identity again, we obtain the action of the Casimir 
\begin{equation}
    \mathcal{C}_{1\ldots j}=\mathcal{C}^{\rm int}\;,
\end{equation}
where 
\begin{equation}
    \mathcal{C}_{1\ldots j}=\frac{1}{2}\left(L^{(1)}_{IJ}+L^{(2)}_{IJ}+\ldots+L^{(j)}_{IJ}\right)\left(L^{(1),IJ}+L^{(2),IJ}+\ldots+L^{(j),IJ}\right)\;,
\end{equation}
\begin{equation}
    \mathcal{C}^{\rm int}=\frac{1}{2}L^{\rm int}_{IJ}L^{{\rm int},IJ}\;.
\end{equation}
However, since $\mathcal{C}^{\rm int}$ acts on the eigenstate we can just replace it by the eigenvalue of the Casimir. This gives us the  R-symmetry Casimir equation which is similar to the conformal Casimir equation for conformal blocks \cite{Dolan:2003hv}.

These Casimir operators can be conveniently used in our setup because the correlator is a polynomial in $y_{ij}$ and the R-symmetry generators simply act as differential operators
\begin{align}
L_{IJ}^{(i)} = y_{i,I}\frac{\partial}{\partial y_{i}^J} -y_{i,J}\frac{\partial}{\partial y_{i}^I}\;. 
\end{align}
For six-point functions, we only encounter two-particle and three-particle Casimirs as nontrivial operators\footnote{The four-particle Casimir is the same as the two-particle Casimir thanks to the invariance of the six-point function.} 
\begin{align}
    \mathcal{C}_{ij} =& \frac{1}{2}\left(L_{IJ}^{(i)}+L_{IJ}^{(j)}\right)\left(L^{(i),IJ}+L^{(j),IJ}\right)\;,\\
     \mathcal{C}_{ijk} =& \frac{1}{2}\left(L_{IJ}^{(i)}+L_{IJ}^{(j)}+L_{IJ}^{(k)}\right)\left(L^{(i),IJ}+L^{(j),IJ}+L^{(k),IJ} \right)\;.
\end{align}
Since the exchanges are associated with single-trace operators, only supergraviton, R-symmetry current and stress tensor can appear. Their eigenvalues for the R-symmetry Casimir are $-24$, $-16$ and $0$ respectively. 

Finding the R-symmetry structures associated with the exchanged representations now becomes a  straightforward task with the use of the R-symmetry Casimirs. We can start with a general polynomial ansatz and then try to solve for the eigensates. Note that we can impose Casimir equations in up to three compatible channels. For example, let us consider simultaneously exchanging single-trace operators in $(12)$, $(34)$ and $(56)$ channels with spins $J_1$, $J_2$ and $J_3$. Since these single-trace operators all have different spins and R-symmetry representations, we can use their spins to label the exchanged operators and denote the corresponding R-symmetry structure by $r_{J_1J_2J_3}$. It is easy to find that for each choice of $\{J_i\}$ there is a unique solution (if the solution exists). Some examples are
\begin{align}
    r_{222} = y_{12}^2y_{34}^2y_{56}^2\;, \ \ \  r_{211} =y_{12}^2y_{34}y_{56}(y_{36}y_{45}-y_{35}y_{46})\;, \label{eq:rsymmetryeigenfunct1}
\end{align}
and there are 13 other eigenstates which are given in Appendix \ref{App:RsymmCaseigen}. Here we have choosen a random normalization because the Casimir equations do not fix the overall factor. We can also impose Casimir equations for only two channels, e.g., in $(12)$ and $(34)$. Apparently, the solutions will not be unique because one can further consider exchanging representations in $(56)$ and there are in general multiple options. Therefore, we will denote the solutions as $r_{J_1J_2;i}$ where the extra label $i$ is introduced to distinguish degenerate solutions. Here we present some solutions   
\begin{align}
    &r_{22} = y_{12}^2y_{34}^2y_{56}^2\;, \ \  \ r_{21} = y_{12}^2 y_{34}y_{56}(y_{36}y_{45}-y_{35}y_{46})\;,\\
    &r_{11;1} = y_{12}y_{34}(y_{13}y_{24}-y_{14}y_{23})y_{56}^2\;, \ \ r_{11;2} = y_{12}y_{34}(y_{16}y_{25}-y_{15}y_{26})(y_{36}y_{45}-y_{35}y_{46})\;,\nonumber\\
    &r_{11;3} = y_{12}y_{34}y_{56}[y_{26}(y_{13}y_{45}-y_{14}y_{35})+y_{16}(y_{24}y_{35}-y_{23}y_{45})]\;,  \nonumber\\
    &r_{11;4} = y_{12}y_{34}y_{56}[y_{25}(y_{13}y_{46}-y_{14}y_{36})+y_{15}(y_{24}y_{36}-y_{23}y_{46})]\;,  \nonumber
\end{align}
while the others structures are not needed since they always involve the exchange of a scalar operator. These contributions can be obtained from the explicit  four-point functions with one spinning and three scalars \cite{Goncalves:2019znr,Goncalves:2023oyx}.

\section{Ansatz and strategy}\label{sec:AnsatzStrategy}
\begin{figure}
 	\centering
 	\includegraphics[width=\linewidth]{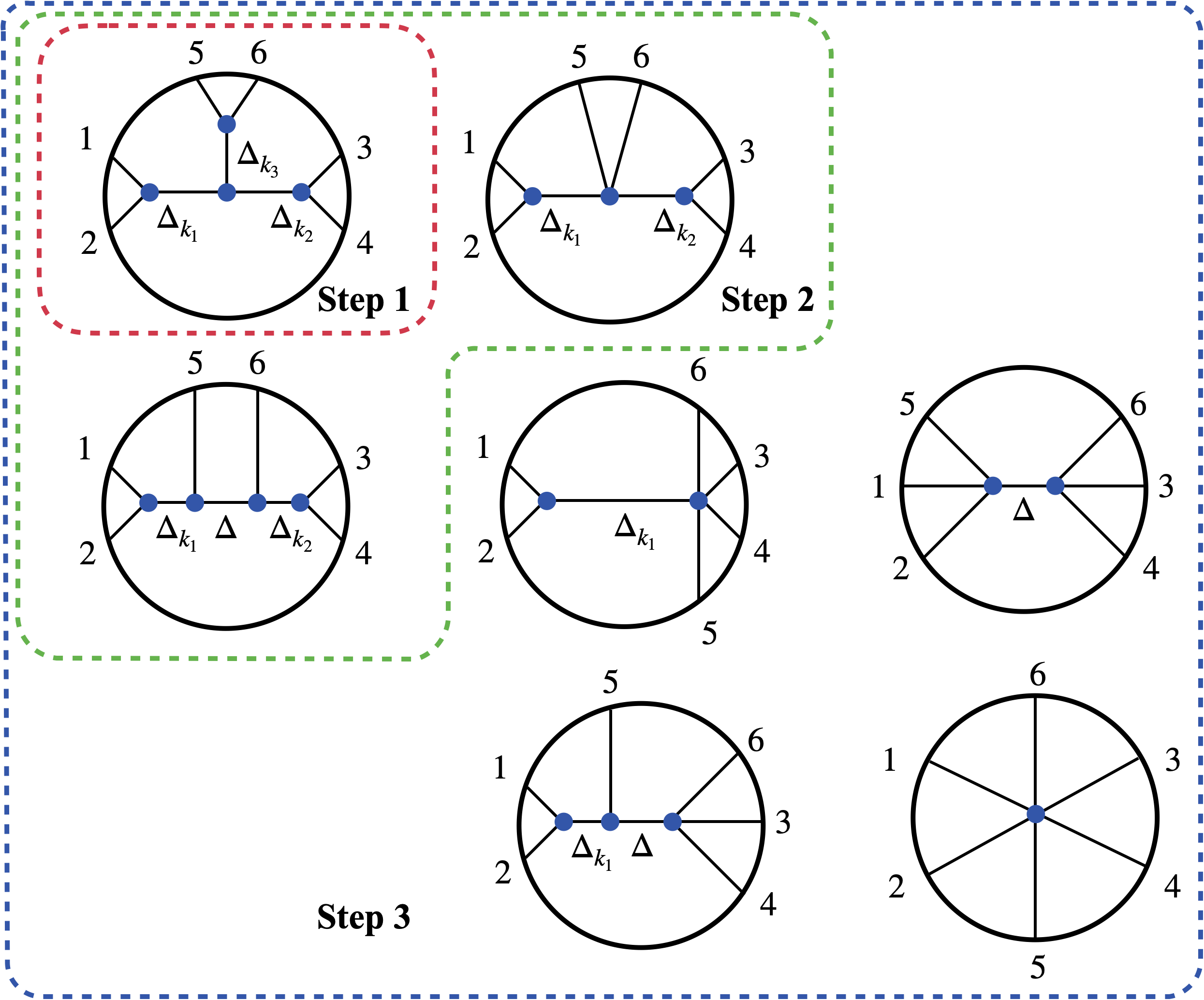}
 	\caption{Illustration of the strategy for computing the supergraviton six-point amplitude. Here we enumerated Witten diagrams of different topology up to permutations and the internal lines can be supergraviton, R-symmetry current and stress tensor. These diagrams are used to represent different parts of the ansatz. Our algorithm is a three-step procedure where at each step an increasingly larger part of the ansatz is solved.}
 	\label{fig:strategy}
 \end{figure}

In this section, we write down the explicit ansatz for the six-point function and outline our strategy. It follows from the properties of the underlying Witten diagrams that the Mellin amplitude of the supergraviton six-point function is a rational function which has poles in $\delta_{ij}$ and is a polynomial in $y_{ij}$. It is convenient to parameterize the ansatz as follows\footnote{This ansatz for supergravitons has a similar structure as the one for supergluons in AdS \cite{Alday:2023kfm}. The main differences are that here $m$ runs up to $2$ instead of $1$ and the degrees of the polynomials in the residues are higher. Moreover, there is a regular term $P_1$ which is absent in the supergluon case.}
\begin{align}
&\mathcal{M}(\delta_{ij},y_{ij}) =\left( \frac{P_4(\delta_{ij},y_{ij})}{(\delta_{12}-1)(\delta_{34}-1)(\delta_{56}-1)}+\textrm{perm}\right)+\left( \frac{P_3(\delta_{ij},y_{ij})}{(\delta_{12}-1)(\delta_{34}-1)}+\textrm{perm}\right)\nonumber\\
&+\left( \frac{P_2(\delta_{ij},y_{ij})}{(\delta_{12}-1)}+\textrm{perm}\right) +\left(\sum_{m=0}^{m_{\textrm{max}=2}} \frac{B_{m}(\delta_{ij},y_{ij})}{(\delta_{12}+\delta_{13}+\delta_{23}+m-2)}+\textrm{perm}\right)+P_{1}(\delta_{ij},y_{ij})\;.\label{eq:ansatz6pt}
\end{align}
Here we have only written one representative term of each type and the whole ansatz involves the sums over all permutations which are denoted by perm. Let us now unpack this ansatz and justify its details. From the pole structures, it is easy to associate $P_i$ and $B_m$ with various exchange (or contact) processes which are enumerated in Figure \ref{fig:strategy}. Then it is clear that the numerators $P_i$ are polynomials in $\delta_{ij}$ while $B_m$ are rational functions and can have poles in the compatible channels at $\delta_{12},\delta_{13},\delta_{23},\delta_{45},\delta_{45},\delta_{46},\delta_{56}=1.$ Recall the high-energy limit of the Mellin amplitude is related to the flat-space amplitude via (\ref{flatspacerelation}) and the flat-space amplitude grows linearly with energy. This implies that the polynomials $P_4$, $P_3$, $P_2$, $P_1$ should respectively have degrees 4, 3, 2 and 1. In the ansatz, we will write them as general polynomials of the corresponding degrees with unfixed coefficients and include all possible R-symmetry structures. For $B_m$, since it is multiplied by a simple pole, we conclude that it should grow quadratically at large energies. Let us also comment on the positions of poles which are associated with the conformal twists of the exchanged operators and their conformal descendants. Here the poles in $\delta_{12}$ etc truncate at 1, i.e., the leading pole, and the poles in $\delta_{12}+\delta_{13}+\delta_{23}$ truncate at $m_{\rm max}=2$. This is a priori not obvious, but can be concluded from factorization using the exact form of spinning three- and four-point functions as we show in Appendix \ref{App:Mellinfactorizationdetails}. Similar pole truncation phenomena was first observed in high-point functions in \cite{Goncalves:2023oyx}.

Our strategy for computing the six-point amplitude is to divide and conquer and is illustrated in Figure \ref{fig:strategy}. We will proceed in three steps. First, we take three simultaneous lightcone limits for (12), (34) and (56). In Mellin space, this corresponds to taking the residue at $\delta_{12}=\delta_{34}=\delta_{56}=1$ and singles out $P_4$ in the ansatz. We will refer to this channel as the snowflake channel and we will use superconformal symmetry to constrain $P_4$. A key observation of this paper which makes this possible is that the snowflake channel analysis forms a closed sector under the chiral algebra condition. This is because in (\ref{chialg}) we are free to choose $\bar{z}_i$.  This freedom allows us to focus on the triple lightcone limit by taking $\bar{z}_{12}=\bar{z}_{34}=\bar{z}_{56}=0$ and study the consquence of chiral algebra with all other contributions in the ansatz turned off. We will perform this analysis in Section \ref{Sec:snowflake} and the upshot is that there are only two constants unfixed in $P_4$. Next we similarly take two simultaneous lightcone limit by setting $\bar{z}_{12}=\bar{z}_{34}=0$ and impose the chiral algebra condition. This weaker condition allows us to also probe $P_3$ and $B_m$. Note that $B_m$ can also be probed by the factorization of the six-point function into two four-point functions. Since all four-point functions with at most one spinning operators are known, $B_m$ are essentially known and will be used as an input. The analysis is carried out in Section \ref{Sec:twosimulpoles} and we find all but a few coefficients are left unfixed. Finally, we impose the Drukker-Plefka twist (\ref{DrukkerPlefkatwist}) in Section \ref{Sec:imposeDPtwist} which completely fixes the ansatz up to an overall constant. The explicit final result can be found in the ancillary Mathematica notebook which we attach to the arXiv submission.

\section{Lightcone limit and chiral algebra: The snowflake channel}\label{Sec:snowflake}
\begin{figure}
 	\centering
 	\includegraphics[width=0.5\linewidth]{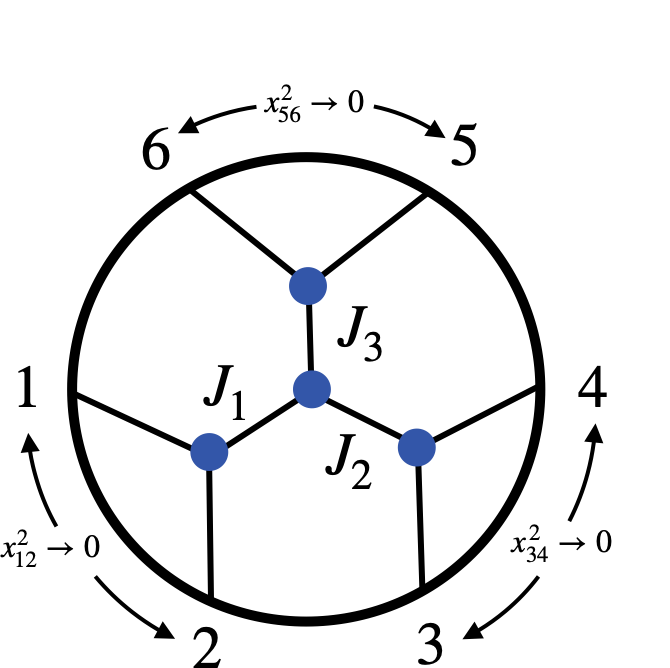}
 	\caption{Snowflake channel.}
 	\label{fig:snowflake}
 \end{figure}

We now implement the strategy outlined in the previous section and start with the snowflake channel (Figure \ref{fig:snowflake}). As explained above, the chiral algebra constraint is compatible with the ligthcone limit, in particular with $x_{12}^2,x_{34}^2,x_{56}^2\to 0$. This allows us to ignore the majority part of the ansatz and study in detail the constraining power of chiral algebra twist entirely within the snowflake channel. 

The Mellin amplitude in the snowflake channel has the following form 
\begin{align}
M(\delta_{ij},y_{ij}) \approx \frac{P_4(\delta_{16},\delta_{23},\delta_{24},\delta_{26},\delta_{45},\delta_{46},y_{ij})}{(\delta_{12}-1)(\delta_{34}-1)(\delta_{56}-1)}+\dots\;,
\end{align}
where $P_4$ is a polynomial of all the other independent Mellin variables and the dots represent regular terms for $\delta_{12}, \delta_{34},\delta_{56}\rightarrow 1$. The flat-space limit (\ref{flatspacerelation}) dictates that $P_4$ has at most degree 4 in the Mellin-Mandelstam variables. Instead of writing down a generic polynomial ansatz, here we take advantage of the fact that the triple factorization is the simultaneous eigenstate of the two-particle conformal Casimirs of 12, 34 and 56. As was explained in Section \ref{Sec:fac3spinning}, this corresponds to the basis $M_{J_1J_2J_3}^{\ell_1\ell_2\ell_3}$ in Mellin space and we can write the ansatz for $P_4$ as   
\begin{align}\label{P4ansatz}
P_4(\delta_{16},\delta_{23},\delta_{24},\delta_{26},\delta_{45},\delta_{46},y_{ij}) = \sum_{J_i=0}^{2}\sum_{\ell_i}r_{J_1J_2J_3}p_{J_1J_2J_3}^{\ell_1\ell_2\ell_3} M_{J_1J_2J_3}^{\ell_1\ell_2\ell_3}(\delta_{16},\delta_{23},\delta_{24},\delta_{26},\delta_{45},\delta_{46})\;,
\end{align}
where the dependence on the polarization vectors $y_I$ is encoded in the R-symmetry eigenstates  $r_{J_1J_2J_3}$  and $p_{J_1J_2J_3}^{\ell_1\ell_2\ell_3}$ are unknown parameters. On the other hand, it should be noted that some $M_{J_1J_2J_3}^{\ell_1\ell_2\ell_3}$ in (\ref{P4ansatz}) have degree higher than 4. Requiring that $P_4$ has only degree 4 already imposes relations among $p_{J_1J_2J_3}^{\ell_1\ell_2\ell_3}$. More precisely when $J_1=J_2=J_3=2$, i.e., gravitons are exchanged in all three channels, there are terms with degrees 5 and 6. Requiring the cancellation of such terms gives us the following relations between different $p_{222}^{\ell_1\ell_2\ell_3}$
\begin{align}
&p_{222}^{101}= \frac{1}{24} \left(661 p_{222}^{000}+125 p_{222}^{001}+9 p_{222}^{002}-38 p_{222}^{010}-8 p_{222}^{011}-18 p_{222}^{020}-96 p_{222}^{100}\right)\;,\\
&p_{222}^{110}= \frac{1}{24} \left(18 p_{222}^{002}+125 p_{222}^{010}+8 p_{222}^{011}-661 p_{222}^{000}-38 p_{222}^{001}-9 p_{222}^{020}+96 p_{222}^{100}\right)\;,\nonumber\\
&p_{222}^{111}= \frac{1}{216} \left(16838 p_{222}^{000}+853 p_{222}^{001}-495 p_{222}^{002}-853 p_{222}^{010}+440 p_{222}^{011}-495 p_{222}^{020}-2448 p_{222}^{100}\right)\;,\nonumber\\
&p_{222}^{200}= \frac{1}{27} \left(661 p_{222}^{000}+125 p_{222}^{001}+9 p_{222}^{002}-125 p_{222}^{010}-32 p_{222}^{011}+9 p_{222}^{020}-9 p_{222}^{100}\right)\;.\nonumber
\end{align}
Looking at the explicit expressions for $M_{J_1J_2J_3}^{\ell_1\ell_2\ell_3}$, the constraint on the degree could also be violated for $J_1=J_2=2$, $J_3=1$. However, such a process is forbidden by R-symmetry. 

To implement the chiral algebra constraint, it is most straightforward to work with position space. The map (\ref{mapGM3pt}) allows us to translate the ansatz (\ref{P4ansatz}) into position space 
\begin{align}\label{G6snowflakelightcone}
&G_{6}(x_{ij}^2,y_{ij}) = \sum_{J_i,\ell_i} \frac{r_{J_1J_2J_3} p_{J_1J_2J_3}^{\ell_1\ell_2\ell_3} }{(x_{12}^2x_{34}^2x_{56}^2)^{ 2}}  G_{ J_1J_2J_3}^{\ell_1\ell_2\ell_3}(u_{i},U_i)+\dots\;,
\end{align}
where $G_{ J_1J_2J_3}^{\ell_1\ell_2\ell_3}(u_{i},U_i)$ is the snowflake conformal block and $p_{J_1J_2J_3}^{\ell_1\ell_2\ell_3}$ is a  product of OPE coefficients. 

Writing $x_{ij}^2=z_{ij}\bar{z}_{ij}$, the chiral algebra twist is implemented by setting $y_{ij}=z_{ij}v_{ij}$ while the lightcone limit is implemented by taking $\bar{z}_{12}=\bar{z}_{34}=\bar{z}_{56}=0$. The chiral algebra constraint is the statement that the twisted correlator is independent of $z_{ij}$. The constraining power comes from the fact that this anti-meromorphicity is not satisfied by individual lightcone conformal blocks. Concretely, we can expand (\ref{G6snowflakelightcone}) in powers of $z_{12}$, $z_{34}$, $z_{56}$ up to the third order \footnote{Alternatively we can also evaluate the conformal block in the planar kinematics for fixed $z_i$. }. This gives linear constraints for $p_{J_1J_2J_3}^{\ell_1\ell_2\ell_3}$ which are solved to give the following relations relating different spins
\begin{align}
&p_{001}^{000}=\frac{1}{2} p_{000}^{000}\;,\quad\quad p_{002}^{000}=-\frac{2}{15}  p_{000}^{000}\;,\quad\quad p_{010}^{000}=\frac{1}{2} p_{000}^{000}\;,\quad\quad p_{011}^{100}=\frac{1}{3} p_{000}^{000}+p_{011}^{000}\;,\nonumber\\
&p_{020}^{000}=-\frac{2}{15}  p_{000}^{000}\;,\quad\quad p_{100}^{000}=-\frac{1}{2} p_{000}^{000}\;,\quad\quad p_{101}^{010}=\frac{2}{9} p_{000}^{000}+p_{101}^{000}\;,\quad\quad p_{110}^{001}=\frac{1}{3} p_{000}^{000}-p_{110}^{000}\;, \nonumber\\
&p_{111}^{100}=\frac{2}{3} p_{000}^{000}+p_{111}^{000}+p_{111}^{001}-p_{111}^{010}\;,\quad\quad p_{112}^{110}=-\frac{8}{27}  p_{000}^{000}-p_{112}^{000}-p_{112}^{001}+p_{112}^{010}+p_{112}^{100}\;,\nonumber\\
&p_{121}^{101}=-\frac{8}{27}  p_{000}^{000}+p_{121}^{000}+p_{121}^{001}-p_{121}^{010}-p_{121}^{100}\;,\label{eq:threepointchiralrelations}\\
&p_{200}^{000}=-\frac{2}{15}  p_{000}^{000}\;,\quad\quad p_{211}^{100}=-\frac{8}{27}  p_{000}^{000}+p_{211}^{000}+p_{211}^{001}-p_{211}^{010}-p_{211}^{011}\;,\nonumber\\
&p_{222}^{200}=-\frac{112}{675} p_{000}^{000}-p_{222}^{000}-p_{222}^{001}-p_{222}^{002}+p_{222}^{010}+p_{222}^{011}-p_{222}^{020}+p_{222}^{100}+p_{222}^{101}-p_{222}^{110}-p_{222}^{111}\;.\nonumber 
\end{align}
We also note that the chiral algebra constraint is not able to distinguish coefficients with different values of $\ell_i$. This is easy to understand because different three-point tensor structures can become degenerate in the two dimensional kinematics. 

To proceed further, we note that there are two more constraints that can be used. The first is the conservation equations for the R-symmetry current and the stress tensor, which allows us to constrain out-of-plane degrees of freedom. Imposing conservation, we find that almost all $p_{J_1J_2J_3}^{\ell_1\ell_2\ell_3}$ are fixed in terms of $p_{000}^{000}$, $p_{111}^{000}$ and $p_{112}^{000}$. In particular, the three-point function coefficients involving $p_{222}^{\ell_1\ell_2\ell_3}$ are completely fixed in terms of $p_{000}^{000}$ and agree with the result for three-point functions of stress tensors \cite{Howe:1998zi,Zhiboedov:2012bm}. The second constraint comes from the fact that in the flat-space amplitude only spin two gravitons can be exchanged since it is a tree-level graviton amplitude. Imposing that the flat-space limit of the Mellin amplitude is dominated by graviton exchanges, we can fix the value of $p_{112}^{000}$ in terms of $p_{000}^{000}$. All in all, the analysis in the snowflake channel tells us that this part of the ansatz is completely determined up to only two unknowns, namely, $p_{000}^{000}$ and $p_{111}^{000}$.

\section{Lightcone limit and chiral algebra: Two simultaneous poles}\label{Sec:twosimulpoles}

\begin{figure}
 	\centering
 	\includegraphics[width=\linewidth]{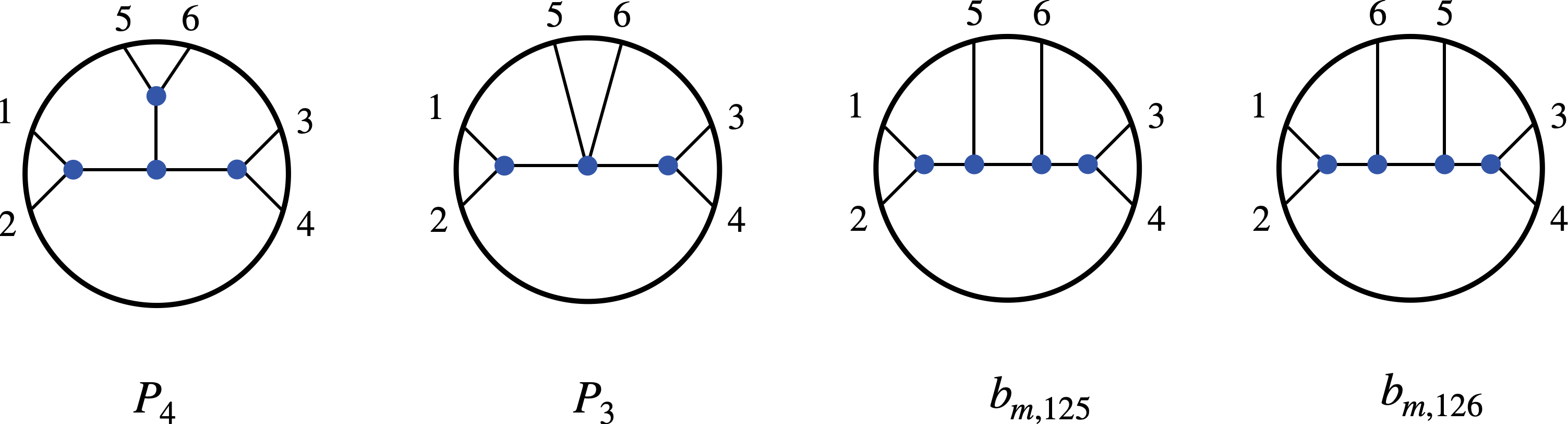}
 	\caption{Nonequivalent processes with simultaneous poles in $\delta_{12}$ and $\delta_{34}$.}
 	\label{fig:simultaneouspoles}
 \end{figure}

In this section we move to a larger subset of the ansatz that has at least two simultaneous poles, say at $\delta_{12}=1$ and $\delta_{34}=1$. More specifically, we will focus on the following terms of (\ref{eq:ansatz6pt}) 
\begin{align}
    &M(\delta_{ij},y_{ij}) = \frac{1}{(\delta_{12}-1)(\delta_{34}-1)}\bigg[\frac{P_4(\delta_{ij},y_{ij})}{(\delta_{56}-1)}+P_3(\delta_{ij},y_{ij})+\sum_{m=0}^{2}\bigg(\frac{b_{m,125}(\delta_{ij},y_{ij})}{(\delta_{12}+\delta_{15}+\delta_{25}+m-2)}\nonumber\\
    &\frac{b_{m,126}(\delta_{ij},y_{ij})}{(\delta_{12}+\delta_{16}+\delta_{26}+m-2)}\bigg)\bigg]+\dots\;.\label{eq:ansatzdoublepole}
\end{align}
In this part, $P_4$ has been fixed in the previous section up to two free coefficients. $P_3$ is an unknown polynomial of degree three in the Mellin variables and is our main target here. The polynomials $b_{m,ijk}$ are the residues of $B_m$ in (\ref{eq:ansatz6pt}) at the poles $\delta_{12}=1$, $\delta_{34}=1$ and $\delta_{12}+\delta_{15}+\delta_{25}+m=2$. As we have mentioned in Section \ref{sec:AnsatzStrategy}, $B_m$ are determined by the known four-point functions via factorization. Therefore, we should treat $b_{m,ijk}$ as input in this sector. The meaning of each term is illustrated in Figure \ref{fig:simultaneouspoles}.

Our strategy to fix this part of the ansatz will be similar. We will apply two lightcone OPEs to single out these terms and then use chiral algebra to constrain them. However, unlike the snowflake channel, taking the simultaneous lightcone OPEs gives us four-point functions instead of three-point functions. Four-point functions depend on two conformal cross ratios and implementing the chiral algebra by simply expanding in $z_{ij}$ would be unfeasible. Instead, a better method is to use the Mellin representation and then impose the chiral algebra constraint in Mellin space. Let us also note that the co-plane condition of chiral algebra leads to no loss of generality here because a four-point function can always be put on a plane by using conformal symmetry.

The implementation of chiral algebra twist can be done in the following way. For each pair of spins $(J_1,J_2)$ exchanged in the channels $(12)(34)$ we need to include all the R-symmetry structures allowed for the respective exchanged operator, as explained in Section \ref{sec:rsymmetry} and more specifically (\ref{eq:twolightconespos}),  
\begin{align}
    \langle \mathcal{O}_{20'} (x_1,y_1)\dots \mathcal{O}_{20'} (x_6,y_6) \rangle   \approx \sum_{J_1J_2,q} \frac{C_{12J_1}C_{34J_2} r_{J_1J_2,q}}{(x_{12}^2 x_{34}^2 )^2} \int [dt_1dt_2]  \bar{G}_{J_1J_2,q}(x_i,t_1,t_2)\;, 
\end{align}
where $\bar{G}$ represents the integrand of (\ref{eq:twolightconespos}) and $q$ labels possible degeneracies for the R-symmetry eigenstates,  $r_{J_1J_2,q}$ , for the Casimir of R-symmetry. Next, we set all points to be on a 2d plane. Since the function $f_{k_1k_2\ell}^{J_1J_2}$ in (\ref{eq:fourpointegrals1}) depends on two four-point conformal cross ratios we do not loose generality in $f_{k_1k_2\ell}^{J_1J_2}$  when we go to this kinematics. As in the previous section, individual contributions depend on both holomorphic and anti-holomorphic variables. Therefore, the chiral algebra condition will impose relations between four-point functions with different spins and different R-symmetry structures. An example of a relation that follows from this analysis, generalizing the equations we obtained for three-point functions (\ref{eq:threepointchiralrelations}), is given by 
\begin{align}
    &f_{110;1}^{11}(u, v) = f_{001,1}^{11}(u,v)+\left(\bar{z}_5-1\right) \bar{z}_5 f_{000,1}^{11}(u,v)+\left(1-\bar{z}_5\right) f_{010,1}^{11}(u,v)+\bar{z}_5 f_{100}^{11}(u,v)\nonumber\\
    &+\frac{\left(2-4 z_5\right) F^{(0,1)}(u,v)}{9 \left(z_5-1\right) z_5^2 \bar{z}_5}+\frac{2 \left(2 z_5-1\right) F^{(1,0)}(u,v)}{9 \left(z_5-1\right) z_5^2 \left(\bar{z}_5-1\right) \bar{z}_5}+\frac{4 F^{(1,1)}(u,v)}{9 z_5^3 \bar{z}_5^2}-\frac{2 \left(\bar{z}_5-1\right) F^{(0,2)}(u,v)}{9 z_5^3 \bar{z}_5^2}\nonumber\\
    &-\frac{2 F^{(2,0)}(u,v)}{9 z_5^3 \left(\bar{z}_5-1\right) \bar{z}_5^2}+\frac{\left(6-4 z_5\right) F(u,v)}{9 \left(z_5-1\right)^2 \left(\bar{z}_5-1\right)},  \ \ \ u=\frac{1}{z_{5}\bar{z}_5}, \ \ v= \frac{(1-z_5)(1-\bar{z}_5)}{z_{5}\bar{z}_5}\;.\label{eq:FourpointchiralRelation}
\end{align}
For convenience, we have used a conformal frame which fixes three points $x_6\rightarrow \infty$, $ x_4=(1,0,\ldots)$, $x_2=(0,0,\ldots)$. The function  $F(u,v)$ is the dynamical part of the four-point function of $20'$ operators
\begin{align}
&\langle {\mathcal{O}}_{20'}(x_2,y_2){\mathcal{O}}_{20'}(x_4,y_4){\mathcal{O}}_{20'}(x_5,y_5) {\mathcal{O}}_{20'}(x_6,y_6) \rangle \nonumber\\ & = \frac{y_{24}^2y_{56}^2(z_5-\alpha_5)(z_5-\bar{\alpha}_5)(\bar{z}_5-\alpha_5)(\bar{z}_5-\bar{\alpha}_5)}{x_{25}^2x_{46}^2 x_{26}^2x_{45}^2(1-z_5)^2(1-\bar{z}_5)^2}\,  F(u,v)\nonumber\;,\\
&u= \frac{x_{24}^2 x_{56}^2}{x_{25}^2 x_{46}^2}=\frac{1}{z_5\bar{z}_5}\;, \ \ v = \frac{x_{26}^2 x_{45}^2}{x_{25}^2 x_{46}^2}=\frac{(1-z_5)(1-\bar{z}_5)}{z_{5}\bar{z}_5} \;, \ \  \\
& \frac{y_{24} y_{56}}{y_{25} y_{46}}=\frac{1}{\alpha_5\bar{\alpha}_5}\;, \ \   \frac{y_{26}y_{45}}{y_{24}y_{56}}=\frac{(1-\alpha_5)(1-\bar{\alpha}_5)}{\alpha_{5}\bar{\alpha}_5} \;.\nonumber
\end{align}
Its derivatives with respect to $u$ and $v$ are denoted as $F^{(i,j)}(u,v)$. We have also used a slightly different expression than (\ref{eq:fourpointegrals1})  for a spinning four-point function
\begin{align}
&\langle \mathcal{O}_{J_1}(x_2,z_2)  \mathcal{O}(x_5)\mathcal{O}(x_6)\mathcal{O}_{J_2}(x_4,z_4)  \rangle=\\
&=\sum_{k_j,\ell,i}\frac{H_{24}^{\ell } V_{2,45}^{k_1} V_{4,25}^{k_2} V_{2,56}^{J_1-k_1-\ell } V_{4,56}^{J_2-k_2-\ell } }{\left(x_{56}^2\right)^{\Delta_{\mathcal{O} }  } \left(x_{24}^2\right)^{\frac{2\tau+J_1+J_2 }{2}  }  } \left(\frac{x_{25}^2}{x_{45}^2}\right)^{\frac{J_2-J_1}{2} } r_{J_1J_2,i}f_{k_1k_2\ell ,i}^{J_1J_2}\left(\frac{x_{24}^2 x_{56}^2}{x_{25}^2 x_{46}^2},\frac{x_{26}^2 x_{45}^2}{x_{25}^2 x_{46}^2}\right)\;,\nonumber
\end{align}
to take into account the R-symmetry eigenfunctions. 

There is an obvious obstruction to translating these relations into Mellin space which is the asymmetric appearance of $z_5$ and $\bar{z}_5$. By contrast, in the Mellin representation, which is defined through $u$ and $v$, $z_5$ and $\bar{z}_5$ appear symmetrically. This problem can be fixed by using the strategy introduced in \cite{Zhou:2017zaw,Alday:2020dtb}, which we review in the following. Let us denote the relation among correlators as
\begin{align}
    W(z_5,\bar{z}_5) = 0\;,
\end{align}
where we highlight the dependence on $z_5$ and $\bar{z}_5$. Although $W(z_5,\bar{z}_5)$ is not symmetric under $z_5\leftrightarrow\bar{z}_5$, the following two combinations are
\begin{equation}
 W_+=W(z_5,\bar{z}_5)+W(\bar{z}_5,z_5)\;,\quad W_-=\frac{W(z_5,\bar{z}_5)-W(\bar{z}_5,z_5)}{z_5-\bar{z}_5}\;.    
\end{equation}
Moreover, one can check that the $z_5$, $\bar{z}_5$ dependence in these combinations can always be written as polynomials of $u$ and $v$. These can be easily translated into Mellin space as difference operators by shifting the Mellin-Madelstam variables. Following this procedure, we obtain difference equations in Mellin space relating different four-point Mellin amplitudes. In addition to relations such as (\ref{eq:FourpointchiralRelation}) which involve new unknown spinning four-point functions, there are also other relations where all four-point are known from previous analyses. As a consistency check, we checked that all these relations hold in Mellin space. 

Implementing the chiral algebra twist, we find that together with conservation equations for spinning operators these conditions are enough to completely fix all the unknown coefficients in the double pole part of the ansatz  (\ref{eq:ansatzdoublepole}).

\section{Imposing the Drukker-Plefka twist}\label{Sec:imposeDPtwist}
The only remaining unknown parameters in the ansatz (\ref{eq:ansatz6pt}) reside in the single poles  and the regular term. Note that $B_m$ are determined by the known four-point functions. Therefore, only $P_2$ and $P_1$ need to be determined. Note that the term with $P_1$ corresponds to the exchange of particles which factorize the six-point amplitude into three-point and five-point amplitudes. Therefore, it should be the sum of these three types of contributions where exchanges are characterized by satisfying the corresponding conformal and R-symmetry Casimir equations. The exchanges of the R-symmetry current and the stress tensors are further constrained by the conservation equations. These constraints leave only $4$ unfixed coefficients in the ansatz for $P_2$ that are present in the contribution of the stress tensor for the single pole\footnote{The reader might wonder why there are only unfixed coefficients in the contribution of the stress tensor. The correlator $\langle\mathcal{O} \mathcal{O} \mathcal{O} \mathcal{O} \mathcal{O} \rangle$ is known. The current correlator $\langle \mathcal{J}\mathcal{O}\mathcal{O}\mathcal{O}\mathcal{O} \rangle$ is largely fixed by the analysis of the previous sections with only the regular term being unknown. This regular term is highly fixed by the fact that it should not contribute to the flat-space limit of six-point correlator. On top of that, the conservation equation  fixes the regular term in terms of the polar part.}. The chiral algebra twist still imposes nontrivial constraints for the single pole sector which can be extracted by taking one lightcone OPE. However, the co-plane configuration is not generic for five-point functions and makes the analysis much more complicated. We will not pursue this analysis because we will find these additional constraints are not needed. Finally, the regular term $P_1$ is a polynomial of degree $1$ and depends on all R-symmetry structures.  This part contributes another $10$ unknown coefficients to the ansatz.

To completely determine the ansatz, the final step of our strategy is to implement the Drukker-Plekfa twist (\ref{DrukkerPlefkatwist}). Note that the condition (\ref{DrukkerPlefkatwist}) is formulated in position space while our ansatz is in Mellin space. Translating the Mellin space ansatz into position space in order to implement the twist would lead to an enormous amount of work. Fortunately, we can also just translate (\ref{DrukkerPlefkatwist}) into Mellin space and this can be easily done. This was first explained in \cite{Goncalves:2023oyx} and we briefly review the translation procedure below. 

The translation of the RHS of (\ref{DrukkerPlefkatwist}) is trivial. According to \cite{Rastelli:2016nze,Rastelli:2017udc}, the Mellin transform of a constant should just be treated as zero if only the Mellin amplitude is concerned. To translate the LHS, we note the following. The Drukker-Plefka twist requires us to keep $x_{ij}^2$ general while setting $y_{ij}=x_{ij}^2$. Since the correlator depends on $y_{ij}$ as a polynomial, a generic monomial becomes a multiplicative $x_{ij}^2$ factor
\begin{equation}
    \prod_{i<j} (y_{ij})^{a_{ij}}\to  \prod_{i<j} (x^2_{ij})^{a_{ij}}\;.
\end{equation}
Comparing with the definition of Mellin amplitude (\ref{eq:MellinAmpDef}), we find that this factor can be absorbed by shifting the Mellin-Mandelstam variables $\delta_{ij}$. This turns each monomial into a difference operator which acts not only on the Mellin amplitude but also on the Gamma factors. Note that in this process we have not kept track of the Mellin integral contours which are also affected by shifting $\delta_{ij}$. The effects from contours are important because as was explained in \cite{Rastelli:2016nze,Rastelli:2017udc} pinching mechanisms can produce nonzero terms in position space from vanishing amplitudes. A more careful analysis taking contours into account should reproduce the constant term on the RHS of (\ref{DrukkerPlefkatwist}). To better understand the implementation, it is instructive to look at an explicit example. Let us focus on
\begin{align}
  \left\{\frac{r_{222}\sum_{\ell_i}p_{222}^{\ell_1\ell_2\ell_3}M_{J_1J_2J_3}^{\ell_1\ell_2\ell_3}(\delta_{ij})}{(\delta_{12}-1)(\delta_{34}-1)(\delta_{56}-1)},\frac{r_{211}\sum_{\ell_i}p_{211}^{\ell_1\ell_2\ell_3}M_{J_1J_2J_3}^{\ell_1\ell_2\ell_3}(\delta_{ij})}{(\delta_{12}-1)(\delta_{34}-1)(\delta_{56}-1)}\right\}\;,\label{eq:exampleDrukkerPlefka1}
\end{align}
which are part of the contributions to the snowflake channel. Using the explicit expressions of $r_{222}$ and $r_{211}$ and following the prescription, the Mellin space version of the Drukker-Plekfa twist reads
\begin{align}
&(\ref{eq:exampleDrukkerPlefka1}) \rightarrow   \delta_{12}\bigg[\delta_{34}\delta_{56}\sum_{\ell_i}p_{222}^{\ell_i}M_{222}^{\ell_i}(\delta_{ij})   \label{MellinafterDPtwist}  \\
&+\delta_{36} \delta_{45}\sum_{\ell_i}p_{211}^{\ell_i}M_{211}^{\ell_i}(\delta_{ij},\delta_{36}+1,\delta_{45}+1)- \delta_{35}\delta_{46}\sum_{\ell_i}p_{211}^{\ell_i}M_{211}^{\ell_i}(\delta_{ij},\delta_{35}+1,\delta_{46}+1)\bigg]\;. \nonumber 
\end{align}
Here the extra factors involving Mellin variables come from the ratio between the shifted Gamma functions and the original ones.  Note that the denominators present in (\ref{eq:exampleDrukkerPlefka1}) are precisely cancelled by this mechanism. This is a consequence of the fact that the R-symmetry polynomials contain at least one power of $y_{12}$, $y_{34}$ and $y_{56}$. This example also shows that the Drukker-Plefka twist in general mixes contributions from the singular and regular parts of the Mellin amplitude. We have also used that the function $M_{J_1J_2J_3}^{\ell_1\ell_2\ell_3}$ does not depend on $\delta_{12},\delta_{34},\delta_{56}$.   Implementing the twist to all terms of the ansatz and requiring the sum to vanish, we find that all remaining parameters are fixed up to an overall constant. 

\section*{Comment on chiral algebra and Drukker-Plefka twist at finite coupling}
In this paper, we studied the implications of the chiral algebra twist and Drukker-Plefka twist focusing on the infinite 't Hooft coupling limit. However, these two conditions also have interesting consequences at finite coupling and these consequences can be extracted by using the same reasoning. Note that at finite coupling, in addition to protected single-trace operators, we can also exchange unprotected single-trace operators which are no longer infinitely heavy as at infinite coupling. 

Let us first consider the constraints from the chiral algebra twist, and again in  the lightcone OPE. We consider an $n$-point function of $\frac{1}{2}$-BPS operators and set their dimensions more generically as $\Delta_{\mathcal{O}}$ (but the same for simplicity). Taking the lightcone OPE limit $x_{12}^2\to 0$, the $n$-point function can be written as
\begin{align}
G_n\to \sum_\tau\sum_{m=0}^{\infty}\frac{C_{12,\tau,J}}{(x_{12}^2)^{\Delta_{\mathcal{O} } -\frac{\tau+2m}{2} }}f_{m,\tau,J} (x_{12}\cdot x_{ij},x_{ij}^2,y_1,\dots, y_n),  \ \ \ \ i,j=2,\dots n \;, \label{eq:chiraldrukkerexplanation}
\end{align}
where $f_{m,\tau,J}$ is related to the $(n-1)$-point function involving the new operator from the OPE. To obtain a meromorphic twisted correlator, all the singular terms need to vanish and this requires the conspiracy of operators appearing in these terms of the OPE.\footnote{The case we considered is special in that both external and internal dimensions are small so that there are no subleading singularities.} In fact, we can understand how they vanish in more detail. An important feature at finite coupling is that the conformal dimensions of unprotected operators are generic and do not overlap with protected operators (or their conformal descendants). Therefore, the cancellation of singularities must happen separately for protected and unprotected operators.\footnote{Here we are assuming the unprotected operators have sufficiently small dimensions so they are manifested in the OPE as singularities. However, since their dimensions are continuously parameterized by the coupling we expect the constraints from chiral algebra can be analytically continued and hold beyond this regime.} For the leading singularity in each tower, the analysis is the same as before. Applying our technology, we obtain differential equations among lower-point correlators. The difference this time is that these correlators involve only unprotected operators residing in the same superconformal multiplet.

The non-overlap of the spectra of protected and unprotected operators also allows us to extract more detailed constraints from the Drukker-Plefka twist. Since the twisted correlator is topological, we can also examine how it is satisfied by taking the lightcone OPE. Again the singularities coming from protected and unprotected operators do not overlap and their cancellations must happen independently. This becomes more manifest in Mellin space. In contrast to (\ref{MellinafterDPtwist}), the pole associated with exchanging an unprotected operator will not be cancelled by the ratio of shifted and original Gamma functions because they only contribute zeros at double-trace locations. Therefore, after translating (\ref{DrukkerPlefkatwist}) into Mellin space we can take its residues at poles associated with the unprotected operators. These gives constraints which are associated with the long operators. 

\section{Comparing with the flat-space limit}\label{sec:FSL}
The most convenient way to obtain the tree-level graviton six-point amplitude in flat-space is via the Kawai-Lewellen-Tye relations \cite{Kawai:1985xq}. Denoting the tree-level color ordered gluon six-point amplitude as $A_{\rm gluon}[123456]$, then the graviton amplitude is given by
\begin{equation}\label{KLT}
\begin{split}
    A_{\rm graviton}={}&-\delta_{12}\delta_{45} A_{\rm gluon}[123456]\left(\delta_{35}A_{\rm gluon}[153462]+(\delta_{34}+\delta_{35})A_{\rm gluon}[154362]\right)\\
    {}&+\mathcal{P}(2,3,4)\;,
\end{split}
\end{equation}
where $\mathcal{P}(2,3,4)$ stands for summing over all permutations of the legs 2, 3 and 4. As we mentioned in Section \ref{Sec:Mellinrepresentation}, the flat-space amplitude we obtain from AdS via (\ref{flatspacerelation}) is in the special kinematic configuration where polarizations are orthogonal to  momenta. This offers considerable simplification as all terms with $p_i\cdot y_j$ drops out. To obtain the flat-space graviton amplitude in this configuration, it is sufficient to use in (\ref{KLT}) the gluon amplitude which is also in the orthogonal configuration \cite{Alday:2023kfm}. The explicit expression for the flat-space graviton amplitude is too lengthy to write down and we include it in the ancillary file. However, certain parts of this amplitude look quite simple and suggestive. For example, the triple pole part of the snowflake channel is
\begin{align}
  A_{\rm graviton}\bigg|_{\delta_{12}=\delta_{34}=\delta_{56}=0}=  \frac{y_{12}^2y_{34}^2y_{56}^2}{\delta_{12}\delta_{34}\delta_{56}}\left(\delta _{16} \delta _{24}+\left(\delta _{26}+\delta _{46}\right) \delta _{24}+\delta _{26} \delta _{45}+\delta _{23} \delta _{46}+\delta _{26} \delta _{46}\right)^2\;.
\end{align}
The terms with just poles in $\delta_{12}$ and $\delta_{34}$ are
\begin{align}
        A_{\rm graviton}\bigg|_{\delta_{12}=\delta_{34}=0}=&   A_{\rm graviton}\bigg|_{\delta_{12}=\delta_{34}=\delta_{56}=0} + \frac{y_{12}^2y_{34}^2y_{56}^2}{\delta_{12}\delta_{34}}\bigg(\frac{\left(\delta _{16} \delta _{24}+\delta _{26} \delta _{24}+\delta _{26} \delta _{45}\right){}^2}{\delta _{16}+\delta _{26}} \nonumber\\
      &-\frac{\left(\delta _{23}+\delta _{24}+\delta _{26}\right){}^2 \delta _{46}^2}{\delta _{16}+\delta _{26}+\delta _{56}}\bigg)\;.
\end{align}
In our computation of the supergraviton six-point Mellin amplitude, we have only made minimal use the flat-space amplitude. The only input from flat-space is that the Mellin amplitude should grow linearly at large energies and the contribution from R-symmetry current needs to be subleading. Therefore, comparing the flat-space limit of our Mellin amplitude against (\ref{KLT}) provides a nontrivial check of the correctness of our result. We have verified that (\ref{KLT}) is precisely reproduced by our Mellin amplitude. But it is also useful to look into the details of this comparison. Recall that in the single pole part $Q_3$ in the ansatz (\ref{eq:ansatz6pt}), we were still left with four free coefficients after imposing the Casimir equation and the conservation conditions. While later we fixed them using the Drukker-Plefka twist, these coefficients are also sensitive to the flat-space limit. That they give rise to the same coefficients shows the extent to which the flat-space limit check is nontrivial. Similarly, in the contact part $Q_6$ the linear term coefficients also survive the flat-space limit and can be fixed either by the flat-space amplitude or the Drukker-Plefka twist. 

While the flat-space limit is only used here as a consistency check, we can  exploit it as nontrivial constraint in future studies now that we have gained confidence for our approach. For example, the information injected by the flat-space amplitude would be very useful when we extend the analysis to Mellin amplitudes with more than six points and study stringy corrections.    

\section{Discussion and outlook}\label{sec:DiscussionOutlook}

In this paper, we computed the supergraviton six-point function in $AdS_5\times S^5$ using a new bootstrap strategy. This strategy allowed us to gain a more detailed understanding of the structure of holographic correlator because we can isolate different parts of the correlators and fix them separately. Our results lead to a number of directions for further explorations. 

First, our result for the supergraviton six-point function should be analyzed further. On the one hand, there is a lot of data encoded in the correlator and it would be useful to extract them. For example, by taking one OPE limit we can obtain spinning five-point functions which can be used as inputs to construct higher-point correlators. We can also obtain correlators with double-trace operators which correspond to scattering processes in AdS involving bound states (see, e.g., \cite{Ceplak:2021wzz,Bissi:2021hjk,Ma:2022ihn,Aprile:2024lwy,Bissi:2024tqf} for recent discussions).  On the other hand, it would be nice to rewrite our result in a more illuminating form which manifests its underlying simplicity. This presumably can be achieved by using differential operators. In particular, it would be interesting to see if the dimensional reduction structures observed at four points \cite{Behan:2021pzk,Alday:2021odx} can also be extended to higher points.

Second, a key ingredient of our approach was the use of the chiral algebra twist which relates correlation functions of different operators belonging to the same supermultiplet. In this paper, we mainly focused on the case where the supermultiplets are $\frac{1}{2}$-BPS and correspond to the supergraviton multiplet. However, this technique can be used more generally and other multiplets can be similarly analyzed. It would be interesting to use this approach to derive relations for correlation functions of other operators, in particular, operators residing in unprotected long multiplets. This method is also useful for deriving superconformal blocks for higher-point functions, either in the co-plane configuration or general kinematics. 

Third, it would be interesting to apply our strategy to bootstrap other correlators. In this paper, we focused on the simplest correlator where the operators have the lowest KK weights. However, we can also study in a similar fashion correlators of massive KK modes. A technical complication is that the intermediate exchanges involve massive fields which are no longer constrained by conservation equations. However, this can be compensated by the flat-space limit which was not exploited very much in fixing the massless correlator. Another interesting application is the  stringy corrections to these correlators where we can also take advantage of the tree-level string amplitude known in flat-space.

Finally, another interesting extension is holographic CFTs with conformal defects. In these theories, we have new observables which are correlation functions of local operators in the presence of defects. These are dual to form factors in AdS which describe the scattering of particles with extended objects. The techniques developed here can also be used in the defect case and will be useful for extending the recent bootstrap results \cite{Barrat:2021yvp,Meneghelli:2022gps,Chen:2023oax,Gimenez-Grau:2023fcy,Chen:2023yvw,Chen:2024orp,Zhou:2024ekb,Alday:2024srr} to higher points.

\section*{Acknowledgments}
We would like to thank Joao Vilas Boas, Bruno Fernandes, Francesco Aprile for illuminating discussions. Centro de Fisica do Porto is partially funded by Fundacao
para a Ciencia e a Tecnologia (FCT) under the grant
UID04650-FCUP.  V.G. is supported
by  Fundacao para a Ciencia e Tecnologia (FCT) under the
grant CEECIND/03356/2022, by  FCT grant 2024.00230.CERN and HORIZON-MSCA-2023-SE-01-101182937-HeI.  X.Z. is supported by the
NSFC Grant No. 12275273, funds from Chinese Academy of Sciences, University of Chinese Academy of Sciences, the Kavli Institute for Theoretical Sciences , and the Xiaomi Foundation.

 \appendix
 \section{Casimir eigenfunctions}
\subsection{Conformal Casimir and lightcone conformal blocks}
In the main text we have explained how the conformal block for six points can be obtained from the lightcone OPE formula
\begin{align}
&\mathcal{O} (x_1)\,\mathcal{O}(x_2) \approx  \sum_{k}C_{12J} \int_{0}^{1}  [dt]\,\frac{\mathcal{O}_{J}(x_1+tx_{21},x_{12}) }{(x_{12}^2)^{\frac{2\Delta_{\mathcal{O}}-\tau}{2}}}
+\dots\;, 
\end{align}
together with the form for spinning three-point functions 
\begin{align}
\langle \mathcal{O}_{J_1}(x_1,z_1)\dots\mathcal{O}_{J_3}(x_3,z_3) \rangle  =\sum_{\ell_i} \frac{C_{J_1J_2J_3}^{\ell_1\ell_2\ell_3}V_{1,23}^{J_1-\ell_2-\ell_3}V_{2,31}^{J_2-\ell_1-\ell_3}V_{3,12}^{J_3-\ell_1-\ell_2}H_{12}^{\ell_3}H_{13}^{\ell_2}H_{23}^{\ell_1}}{(x_{12}^2)^{\frac{h_1+h_2-h_3}{2}}(x_{13}^2)^{\frac{h_1+h_3-h_2}{2}}(x_{23}^2)^{\frac{h_2+h_3-h_1}{2}}}\,.\label{eq:SpinningThreepointFunction}
\end{align}
Here we used a null polarization vector $z_i$ to encode the indices of the operators, $h_i=\Delta_i+J_i$ and $V$ and $H$ are defined as
\begin{align}
V_{i,jk} =\frac{ (z_i\cdot x_{ij} ) x_{ik}^2 - (z_i\cdot x_{ik}) x_{ij}^2}{x_{jk}^2} \,, \qquad  H_{ij} =(z_i\cdot x_{ij} )(z_j\cdot x_{ij}) -\frac{x_{ij}^2 (z_i\cdot z_{j} )}{2}\,.
\label{eq:estruturas3pf}
\end{align}
After a simple manipulation of the integrals, which was carefully explained in \cite{Antunes:2021kmm},  it is possible to write the expression for the lightcone conformal block in terms of the cross ratios
\begin{align}
&g_{J_1J_2J_3}^{\ell_1\ell_2\ell_3}(u_{2i},U_i)=\int \prod_{i=1}^3\frac{[dt_i] (U_i-u_{2(2-i)})^{\ell_{1-i}}  U_{i}^{\frac{J_{2i} +J_{2(i-1)} - J_{2i-1} +\Delta_{2i} +\Delta_{2i-1} -\Delta_{2(i-1)}}{2}} A_i^{J_i+\ell_i-L} }{2^{J_i} B_{i}^{\ell_i -\Delta_{i} -L+\frac{\sum_j h_j}{2} } }, \nonumber \\
&A_i= U_{2-i} \left(t_{i-1} \left(\left(u_{2 (i-1)}-1\right) U_{1-i}-U_{3-i}+1\right)-\left(u_{2 (i-1)}-1\right) U_{1-i}\right)\label{eq:conformalblocklightcone}\\
&+t_{i+1} \left(u_{2 i}-U_{2-i}\right) \left(\left(t_{i-1}-1\right) u_{2 (i-1)} U_{1-i}-t_{i-1} U_{3-i}\right)\nonumber\\
&B_i=t_{i+1} U_{1-i}-\left(t_{i+1}-1\right) U_{2-i} \left(t_{i-1} \left(u_{2 (i+1)}-U_{1-i}\right)+U_{1-i}\right), \ \ L=\sum_{j}\ell_j\;,\nonumber
\end{align}
where the indices of the variables $t_i,J_i,\Delta_{i}$ and $U_i$ should be evaluated with $\textrm{mod} \,3$ and the index of $u_{i}$ with $\textrm{mod} \,6$. 

Let us note that the basis of three-point functions that we have used is just a choice. From this point of view it is easy to see that there can be more than one solution for the Casimir equation, for a given set of spins $(J_1,J_2,J_3)$,  which is related to the different tensor structures present in three-point functions.  The Casimir equation is not able to distinguish different values of $\ell_i$, since this is related with a choice of the basis. 

In the main text we have explained how to obtain the Mellin amplitude amplitude of the lightcone conformal block. Here we will write the Mellin transform for all cases but only full expression for two sets $J_i,\ell_i$
\begin{align}
&G_{ J_1J_2J_3}^{\ell_1\ell_2\ell_3}(u_{i},U_i)\;\rightarrow \frac{M_{J_1J_2J_3}^{\ell_1\ell_2\ell_3}}{(\delta_{12}-1)(\delta_{34}-1)(\delta_{56}-1)}\;,\\
&M_{222}^{111} =  \frac{3375}{64}\left(\left(\bar{\delta}_{234}\right) \delta _{26}-\left(\bar{\delta}_{134} \right) \delta _{16}\right) \left(\left(\bar{\delta}_{612}\right) \delta _{46}-\left(\bar{\delta}_{512}\right) \delta _{45}\right) \nonumber\\
&\left(\delta _{24} \left(\bar{\delta}_{456}\right)-\delta _{23} \left(\bar{\delta}_{356}\right)\right)\;,\\
&M_{222}^{110} = \frac{375}{64} \left(\left(\bar{\delta}_{234}\right) \delta _{26}-\left(\bar{\delta}_{134}\right) \delta _{16}\right) \left(\left(\bar{\delta}_{612}\right) \delta _{46}-\left(\bar{\delta}_{512}\right) \delta _{45}\right) \nonumber\\
&\left(23 \delta _{24} \left(\bar{\delta}_{456}\right)-8 \left(2 \delta _{24}+2 \delta _{45}+2 \delta _{46}-1\right)+\delta _{23} \left(23\bar{\delta}_{456}-7\right)\right)\;,
\end{align}
where $\bar{\delta}_{ijk}=\delta_{ij}+\delta_{ik}$.  Writing the complete expressions for all $M_{J_1J_2J_3}^{\ell_1\ell_2\ell_3}$ relevant to this paper would take too much space. For this reason we only quote here the leading term in the flat-space limit of each polynomial $M_{J_1J_2J_3}^{\ell_1\ell_2\ell_3}$. This can be used to check the normalization and to generate the complete function using the Casimir equation in Mellin space. 
\begin{align}
&M_{000}^{000}= 1, \ M_{001}^{000}= 6 \bar{\delta}_{612}, \ M_{002}^{000}= \frac{45}{2} \bar{\delta}_{612}^2, \ M_{010}^{000}= -6 \bar{\delta}_{456}, \ M_{011}^{000}= -\frac{63}{2}\bar{\delta}_{612}\bar{\delta}_{456}, \nonumber\\
&M_{011}^{100}= \frac{9}{2}\bar{\delta}_{612}\bar{\delta}_{456}, \ M_{012}^{000}= -105 \bar{\delta}_{612}^2 \bar{\delta}_{456}, \ M_{012}^{100}= 15 \bar{\delta}_{612}^2 \bar{\delta}_{456}, \ M_{020}^{000}= \frac{45}{2} \bar{\delta}_{456}^2,  \nonumber\\
&M_{110}^{001}= \frac{9}{2} \bar{\delta}_{234}\bar{\delta}_{456},\ M_{021}^{000}= 105 \bar{\delta}_{612}\bar{\delta}_{456}^2, M_{021}^{100}= -15 \bar{\delta}_{612}\bar{\delta}_{456}^2, \ M_{101}^{000}= -\frac{63}{2}\bar{\delta}_{234}\bar{\delta}_{612}, \nonumber\\ 
&M_{022}^{000}= \frac{9825}{32} \bar{\delta}_{612}^2 \bar{\delta}_{456}^2, \ M_{022}^{100}= -\frac{1575}{32}  \bar{\delta}_{612}^2 \bar{\delta}_{456}^2, \ M_{022}^{200}= \frac{75}{32} \bar{\delta}_{612}^2 \bar{\delta}_{456}^2, \ M_{100}^{000}= -6 \bar{\delta}_{234},\nonumber\\
&M_{101}^{010}= \frac{9}{2} \bar{\delta}_{234}\bar{\delta}_{612}, \ M_{102}^{000}= -105 \bar{\delta}_{234}\bar{\delta}_{612}^2, \ M_{102}^{010}= 15 \bar{\delta}_{234}\bar{\delta}_{612}^2, \ M_{110}^{000}= \frac{63}{2} \bar{\delta}_{234}\bar{\delta}_{456}, \nonumber\\
&M_{112}^{001}= \frac{405}{4} \bar{\delta}_{234}\bar{\delta}_{612}^2 \bar{\delta}_{456},\ M_{112}^{000}= \frac{765}{2} \bar{\delta}_{234}\bar{\delta}_{612}^2 \bar{\delta}_{456}, \ M_{111}^{000}= 135 \bar{\delta}_{234}\bar{\delta}_{612}\bar{\delta}_{456}, \  \nonumber\\
&M_{111}^{001}= 27 \bar{\delta}_{234}\bar{\delta}_{612}\bar{\delta}_{456}, \ M_{111}^{010}= -27 \bar{\delta}_{234}\bar{\delta}_{612}\bar{\delta}_{456}, \ M_{111}^{100}= -27 \bar{\delta}_{234}\bar{\delta}_{612}\bar{\delta}_{456}, \nonumber\\
&M_{112}^{010}= -\frac{585}{8}  \bar{\delta}_{234}\bar{\delta}_{612}^2 \bar{\delta}_{456}, \ M_{112}^{100}= -\frac{585}{8}  \bar{\delta}_{234}\bar{\delta}_{612}^2 \bar{\delta}_{456}, \ M_{112}^{110}= \frac{135}{8} \bar{\delta}_{234}\bar{\delta}_{612}^2 \bar{\delta}_{456}, \nonumber\\
&M_{120}^{001}= -15 \bar{\delta}_{234}\bar{\delta}_{456}^2,\, M_{121}^{000}= -\frac{765}{2} \bar{\delta}_{234}\bar{\delta}_{612}\bar{\delta}_{456}^2,\, M_{121}^{001}= -\frac{585}{8}  \bar{\delta}_{234}\bar{\delta}_{612}\bar{\delta}_{456}^2, \ M_{121}^{010}= \frac{405}{4} \bar{\delta}_{234}\bar{\delta}_{612}\bar{\delta}_{456}^2,\nonumber\\
&M_{121}^{101}= \frac{135}{8} \bar{\delta}_{234}\bar{\delta}_{612}\bar{\delta}_{456}^2, \ M_{121}^{100}= \frac{585}{8} \bar{\delta}_{234}\bar{\delta}_{612}\bar{\delta}_{456}^2, \ M_{120}^{000}= -105 \bar{\delta}_{234}\bar{\delta}_{456}^2,\nonumber\\
&M_{122}^{000}= -\frac{15075}{16}  \bar{\delta}_{234}\bar{\delta}_{612}^2 \bar{\delta}_{456}^2, \ M_{122}^{001}= -225 \bar{\delta}_{234}\bar{\delta}_{612}^2 \bar{\delta}_{456}^2, \ M_{122}^{010}= 225 \bar{\delta}_{234}\bar{\delta}_{612}^2 \bar{\delta}_{456}^2,\nonumber\\
&M_{122}^{101}= \frac{225}{4} \bar{\delta}_{234}\bar{\delta}_{612}^2 \bar{\delta}_{456}^2, \ M_{122}^{110}= -\frac{225}{4}\bar{\delta}_{234}\bar{\delta}_{612}^2 \bar{\delta}_{456}^2, \ M_{122}^{200}= -\frac{225}{16} \bar{\delta}_{234}\bar{\delta}_{612}^2 \bar{\delta}_{456}^2\nonumber\\
&M_{202}^{020}= \frac{75}{32} \bar{\delta}_{234}^2 \bar{\delta}_{612}^2,\ M_{122}^{100}= \frac{2925}{16} \bar{\delta}_{234}\bar{\delta}_{612}^2 \bar{\delta}_{456}^2, \ M_{200}^{000}= \frac{45}{2} \bar{\delta}_{234}^2, \ M_{201}^{000}= 105 \bar{\delta}_{234}^2 \bar{\delta}_{612}, \  \nonumber \\
&M_{212}^{010}= \frac{2925}{16} \bar{\delta}_{234}^2 \bar{\delta}_{612}^2 \bar{\delta}_{456}, \ M_{201}^{010}= -15 \bar{\delta}_{234}^2 \bar{\delta}_{612}, \ M_{202}^{000}= \frac{9825}{32} \bar{\delta}_{234}^2 \bar{\delta}_{612}^2, \ M_{202}^{010}= -\frac{1575}{32}  \bar{\delta}_{234}^2 \bar{\delta}_{612}^2,\nonumber\\
&M_{210}^{000}= -105 \bar{\delta}_{234}^2 \bar{\delta}_{456}, \ M_{210}^{001}= -15 \bar{\delta}_{234}^2 \bar{\delta}_{456}, \ M_{211}^{000}= -\frac{765}{2} \bar{\delta}_{234}^2 \bar{\delta}_{612}\bar{\delta}_{456}, \ M_{211}^{001}= -\frac{585}{8}  \bar{\delta}_{234}^2 \bar{\delta}_{612}\bar{\delta}_{456},\nonumber\\
&M_{211}^{100}= \frac{405}{4} \bar{\delta}_{234}^2 \bar{\delta}_{612}\bar{\delta}_{456}, \ M_{212}^{000}= -\frac{15075}{16}  \bar{\delta}_{234}^2 \bar{\delta}_{612}^2 \bar{\delta}_{456}, \ M_{212}^{001}= -225 \bar{\delta}_{234}^2 \bar{\delta}_{612}^2 \bar{\delta}_{456},\nonumber\\
&M_{211}^{011}= \frac{135}{8} \bar{\delta}_{234}^2 \bar{\delta}_{612}\bar{\delta}_{456},\ M_{220}^{001}= \frac{1575}{32} \bar{\delta}_{234}^2 \bar{\delta}_{456}^2, \ M_{212}^{011}= \frac{225}{4} \bar{\delta}_{234}^2 \bar{\delta}_{612}^2 \bar{\delta}_{456}, \nonumber\\
&M_{211}^{010}= \frac{585}{8} \bar{\delta}_{234}^2 \bar{\delta}_{612}\bar{\delta}_{456}, \ M_{212}^{100}= 225 \bar{\delta}_{234}^2 \bar{\delta}_{612}^2 \bar{\delta}_{456}, \ M_{212}^{110}= -\frac{225}{4}\bar{\delta}_{234}^2 \bar{\delta}_{612}^2 \bar{\delta}_{456}, \nonumber\\
&M_{212}^{020}= -\frac{225}{16}\bar{\delta}_{234}^2 \bar{\delta}_{612}^2 \bar{\delta}_{456}, \ M_{220}^{002}= \frac{75}{32} \bar{\delta}_{234}^2 \bar{\delta}_{456}^2,\, M_{221}^{000}= \frac{15075}{16} \bar{\delta}_{234}^2 \bar{\delta}_{612}\bar{\delta}_{456}^2,\, M_{221}^{001}= \frac{2925}{16} \bar{\delta}_{234}^2 \bar{\delta}_{612}\bar{\delta}_{456}^2, \nonumber\\
&M_{221}^{011}= -\frac{225}{4}\bar{\delta}_{234}^2 \bar{\delta}_{612}\bar{\delta}_{456}^2, \ M_{221}^{100}= -225 \bar{\delta}_{234}^2 \bar{\delta}_{612}\bar{\delta}_{456}^2, \ M_{221}^{101}= -\frac{225}{4}\bar{\delta}_{234}^2 \bar{\delta}_{612}\bar{\delta}_{456}^2,
\nonumber\\
&M_{221}^{010}= -225 \bar{\delta}_{234}^2 \bar{\delta}_{612}\bar{\delta}_{456}^2, \ M_{220}^{000}= \frac{9825}{32} \bar{\delta}_{234}^2 \bar{\delta}_{456}^2,\ M_{221}^{002}= \frac{225}{16} \bar{\delta}_{234}^2 \bar{\delta}_{612}\bar{\delta}_{456}^2, \nonumber\\
&M_{222}^{001}= \frac{29625}{64} \bar{\delta}_{234}^2 \bar{\delta}_{612}^2 \bar{\delta}_{456}^2, \ M_{222}^{002}= \frac{3375}{64} \bar{\delta}_{234}^2 \bar{\delta}_{612}^2 \bar{\delta}_{456}^2, \ M_{222}^{010}= -\frac{29625}{64}  \bar{\delta}_{234}^2 \bar{\delta}_{612}^2 \bar{\delta}_{456}^2, \nonumber\\
& M_{222}^{011}= -\frac{8625}{64}\bar{\delta}_{234}^2 \bar{\delta}_{612}^2 \bar{\delta}_{456}^2, \ M_{222}^{020}= \frac{3375}{64} \bar{\delta}_{234}^2 \bar{\delta}_{612}^2 \bar{\delta}_{456}^2, \ M_{222}^{100}= -\frac{29625}{64}  \bar{\delta}_{234}^2 \bar{\delta}_{612}^2 \bar{\delta}_{456}^2, \nonumber\\
&M_{222}^{000}= \frac{129375}{64} \bar{\delta}_{234}^2 \bar{\delta}_{612}^2 \bar{\delta}_{456}^2, \ M_{222}^{101}= -\frac{8625}{64}\bar{\delta}_{234}^2 \bar{\delta}_{612}^2 \bar{\delta}_{456}^2, \ M_{222}^{200}= \frac{3375}{64} \bar{\delta}_{234}^2 \bar{\delta}_{612}^2 \bar{\delta}_{456}^2\;.
\end{align}
\subsection{R-symmetry Casimir eigenfunctions}\label{App:RsymmCaseigen}
In the main text we spelled out the eigenfunctions of the Casimir equation for R-symmetry in three and two non-consecutive channels. In this subsection we collect the remaining eigenfunctions. For the triple Casimir we have the following (up to permutations of the pairs $(12)(34)(56)$)
\begin{align}
&r_{000}= \frac{1}{6} \big(6 y_{34} y_{36} y_{45} y_{56} y_{12}^2+6y_{34} y_{35} y_{46} y_{56} y_{12}^2+6 y_{14} y_{23} y_{34}y_{56}^2 y_{12}+6 y_{13} y_{24} y_{34} y_{56}^2y_{12}\nonumber\\
&+6 y_{16} y_{25} y_{34}^2 y_{56} y_{12}+6 y_{15} y_{26} y_{34}^2 y_{56} y_{12}-9 y_{16} y_{24} y_{34} y_{35} y_{56} y_{12}-9 y_{14} y_{26} y_{34} y_{35} y_{56} y_{12}\nonumber\\
&-9 y_{15} y_{24} y_{34} y_{36} y_{56} y_{12}-9 y_{14} y_{25} y_{34} y_{36} y_{56} y_{12}-9 y_{16} y_{23} y_{34} y_{45} y_{56} y_{12}-9 y_{13} y_{26} y_{34} y_{45} y_{56} y_{12}\nonumber\\
&-9 y_{15} y_{23} y_{34} y_{46}y_{56} y_{12}-9 y_{13} y_{25}y_{34}y_{46}y_{56}y_{12}-4 y_{34}^2 y_{56}^2 y_{12}^2\big),\nonumber\\
&r_{001}= y_{12} y_{16} y_{24} y_{34} y_{35} y_{56}+y_{12} y_{14} y_{26} y_{34} y_{35} y_{56}-y_{12} y_{15} y_{24} y_{34} y_{36} y_{56}-y_{12} y_{14} y_{25} y_{34} y_{36} y_{56}\nonumber\\
&+y_{12} y_{16} y_{23} y_{34} y_{45} y_{56}+y_{12} y_{13} y_{26} y_{34} y_{45} y_{56}-y_{12} y_{15} y_{23} y_{34} y_{46} y_{56}-y_{12} y_{13} y_{25} y_{34} y_{46} y_{56}, \nonumber\\
&r_{002}= \frac{1}{3} \left(3 y_{12} y_{14} y_{23} y_{34} +3 y_{12} y_{13} y_{24} y_{34} -y_{12}^2 y_{34}^2\right)y_{56}^2, \nonumber\\
&r_{011}= \frac{1}{3} \big(2 y_{34} y_{36} y_{45} y_{56} y_{12}^2-2 y_{34} y_{35} y_{46} y_{56} y_{12}^2+3 y_{16} y_{24} y_{34} y_{35} y_{56} y_{12}+3 y_{14} y_{26} y_{34} y_{35} y_{56} y_{12}\nonumber\\
&-3 y_{15} y_{24} y_{34} y_{36} y_{56} y_{12}-3 y_{14} y_{25} y_{34} y_{36} y_{56} y_{12}-3 y_{16} y_{23} y_{34} y_{45} y_{56} y_{12}-3 y_{13} y_{26} y_{34} y_{45} y_{56} y_{12}\nonumber\\
&+3 y_{15} y_{23} y_{34} y_{46} y_{56} y_{12}+3 y_{13} y_{25} y_{34} y_{46} y_{56} y_{12}\big), \nonumber\\
&r_{111}= y_{12} y_{16} y_{24} y_{34} y_{35} y_{56}-y_{12} y_{14} y_{26} y_{34} y_{35} y_{56}-y_{12} y_{15} y_{24} y_{34} y_{36} y_{56}+y_{12} y_{14} y_{25} y_{34} y_{36} y_{56}\nonumber\\
&-y_{12} y_{16} y_{23} y_{34} y_{45} y_{56}+y_{12} y_{13} y_{26} y_{34} y_{45} y_{56}+y_{12} y_{15} y_{23} y_{34} y_{46} y_{56}-y_{12} y_{13} y_{25} y_{34} y_{46} y_{56}, \nonumber
\end{align}

\subsection{Details of Mellin factorization}\label{App:Mellinfactorizationdetails}
A key ingredient in our approach is the use of known lower point functions as input to the ansatz. In the main text we have explained the main ideas of factorization of Mellin amplitudes that were worked out in detail in \cite{Goncalves:2014rfa}. In this appendix we will review how they are used in the present context and highlight the truncation of the poles mentioned in the main text. The scalar exchange has the simplest factorization formula
\begin{align}
    Q_{m} =\frac{-2\Gamma(\Delta) m!}{ 
 \left(1+\Delta-\frac{d}{2}  \right)_m }
 L_m R_m\;, \label{eq:QmScalar1}
\end{align}
where  
\begin{align}
&L_m =\sum_{n_{ab}\ge 0 \atop \sum n_{ab}=m} 
M_L(\delta_{ab}+n_{ab})  
\prod_{1\le a<b\le k} \frac{ 
\left(\delta_{ab}\right)_{n_{ab}}}{n_{ab}!}\label{eq:Lmdef1}\;,
\end{align}
and similarly for $R_m$. Here $M_L$ and $M_R$ stand for the lower point Mellin amplitudes which in our case is the four point function of $20'$ operators. For a constant three-point Mellin amplitude, say $M_L=C$,  and  for the $3-5$ factorization there is a simplification that prevents the value of $m$ to go beyond $0$. We find 
\begin{align}
  L_m = C\frac{ 
\left(\delta_{12}\right)_{m}}{m!} =  C\frac{ 
\left(\frac{\Delta_1+\Delta_2-\tau-2m}{2}\right)_{m}}{m!} =C\frac{ 
\left(1-m\right)_{m}}{m!} \ ,
\end{align}
where we have assumed 1 and 2 are involved in the left part and used the value of $\delta_{12}$ at the pole. The Pochhammer symbol ensures that $L_m$ vanishes for $m>0$. Thus the gluing for this particular channel is particularly simple. We just need to use the known expression for the five-point function \cite{Goncalves:2019znr} and the R-symmetry gluing uses the formula
\begin{align}
y_{1i}y_{2i}\times y_{rj}y_{rk} \rightarrow y_{2j}^{2}y_{1k} + y_{1j}^{2}y_{2k} - \frac{y_{12}y_{jk}}{3}.
\end{align}
The $4-4$ factorization  is more complicated since the lower point Mellin amplitudes depend on the Mellin variables
\begin{align}
&M_{L}(\delta_{ij}) = 2 \left(\frac{2 \delta _{12}-2 \delta _{23}+2}{\delta _{13}-1}-4 \delta _{12}+\frac{2 \delta _{12}-2 \delta _{13}+2}{\delta _{23}-1}-4\right) y_{01} y_{02} y_{13} y_{23}+\\
&+2 \left(\frac{\delta _{12}^2-\delta _{12}+\delta _{23}^2-\delta _{23}}{\delta _{13}-1}+2-\delta _{13}\right) y_{02}^2 y_{13}^2+\dots\;,\nonumber
\end{align}
where $\dots$ denote permutations and we have chosen the lower-point function to depend on the points $1230$ (with $0$ being the point that is glued). The corresponding $L_m$ for this Mellin amplitude is given by
\begin{align}
&L_m=\frac{y_{02} y_{13}}{m!\Gamma (3-m)}\bigg[ 4 y_{01}  y_{23} \bigg(2 \left(m-2 \left(\delta _{12}+1\right)\right)+\frac{(m-2) \left(\delta _{23}-\delta _{12}+m-1\right)}{\delta _{13}-1}\\
&+\frac{(m-2) \left(\delta _{13}-\delta _{12}+m-1\right)}{\delta _{23}-1}\bigg)+2 y_{02} y_{13}\bigg(\frac{2 \left(\delta _{23}^2+(m-1) \delta _{23}+\delta _{12} \left(\delta _{12}+m-1\right)\right)}{\delta _{13}-1}\nonumber\\
&-2 \left(\delta _{13}+m-2\right)\bigg)\bigg]+\dots\; .\nonumber
\end{align}
Note that the Gamma function in the denominator prevents $m$ from going beyond $2$. Experimentally we have verified that the level of truncation is sensitive to the structure of the Mellin amplitude. 
For example, if the Mellin amplitude is a rational function then the truncation will depend  on the location of the singular terms and the degree of polynomials appearing in the residues  (which is related with the degree of the exchanged operator in a four-point function). The residue $Q_m$ is built from the expression above by multiplying $L_m$ with $R_m$ which is obtained by doing the permutation $(123)\rightarrow (456)$. We should also emphasize that the stress tensor is only exchanged in such four-point function where all operators are scalars, as is dictated by R-symmetry selections.  

The factorization formulas for spinning operators are slightly more complicated. For the exchange of the conserved current we have
\begin{align}
 Q_{m}= \frac{-2\Gamma(\Delta) m!}{ 
 \left(1+\Delta-\frac{d}{2}  \right)_m }
\sum_{a=1}^{k}\sum_{b=k+1}^n \delta_{ab} L_m^{a} R_m^{b}\;,
\end{align}
while the formula for the exchange of a conserved spin $2$ operator is given by
\begin{align}
&\mathcal{Q}_m = \frac{m!}{\left(\frac{d}{2} +1\right)_m}\bigg[\mathcal{Q}_{m}^{(1)} -\left(\frac{1}{2m}+\frac{1}{d}\tilde{L}_m \tilde{R}_m  \right)\bigg]\label{eq:Lmtil} \ , \\
&\mathcal{Q}_{m}^{(1)}  = \sum_{a,b=1}^{k}\sum_{i,j=k+1}^{n}\delta_{ai}(\delta_{bj}+\delta_{b}^{a}\delta_{j}^{i} )L_m^{ab}R_{m}^{ij},\, \ \ \ \ \ \tilde{L}_m = \sum_{a,b=1}^{k}\delta_{ab} [ L_{m-1}^{ab} ]^{ab}\;, \nonumber
\end{align}
where we used the notation\footnote{We have used $\delta$ with upper and subscript as a Kronecker delta while other $\delta$'s should be interpreted as Mellin variables. We hope the reader does not get confused with this notation. } $[f(\delta_{ij})]^{ab} = f(\delta_{ij}+\delta_{i}^{a}\delta_{j}^{b}+\delta_{j}^{a}\delta_{i}^{b})$.  The four-point function with one current and three scalars is given by
\begin{align}
&M^{2}_L(\delta) = -\frac{2 y_{12} \left(\left(\delta _{23}+1-\delta _{13}\right) y_{13} Y_{023}+y_{23} (\left(\delta _{23}-\delta_{13}-1\right) Y_{031}\right)}{\delta _{12}-1}\\
&-\frac{2 y_{13} \left(\left(\delta _{12}-\delta _{23}+1\right) y_{23} Y_{021}+\left(\delta _{23}+1-\delta _{12}\right) y_{12} Y_{023}\right)}{\delta _{13}-1}\nonumber\\
&-\frac{2 y_{23} \left(\left(\delta _{12}-\delta _{13}+1\right) y_{13} Y_{021}+y_{12} (\left(\delta _{12}-1-\delta _{13}\right) Y_{013} \right)}{\delta _{23}-1}+4 y_{12} y_{13} Y_{023}+4 y_{23}y_{13} Y_{021}\;,\nonumber
\end{align}
where we have just written down one component (the others can be obtained by symmetry) and $Y_{0ij}$ is an R-symmetry structure associated with the exchange of a current. It is simple to obtain $L_m^{2}$ from the previous expression
\begin{align}
&L_m^{2} = \frac{1}{m!\Gamma(2-m)}  \bigg[ \frac{2 y_{23} \left(y_{13} Y_{021} \left(\delta _{13}-\delta _{12}+m-1\right)-y_{12} Y_{013} \left(\delta _{12}-\delta _{13}+m-1\right)\right)}{\delta _{23}-1}\nonumber\\
&+\frac{2 y_{13} \left(y_{23} Y_{021} \left(\delta _{23}-\delta _{12}+m-1\right)+y_{12} Y_{023} \left(\delta _{12}-\delta _{23}+m-1\right)\right)}{\delta _{13}-1}\\
&+\frac{2 y_{12} \left(y_{23} Y_{013} \left(\delta _{23}-\delta _{13}+m-1\right)+y_{13} Y_{023} \left(\delta _{13}-\delta _{23}+m-1\right)\right)}{\delta _{12}-1}+4 y_{13} \left(y_{23} Y_{021}+y_{12} Y_{023}\right)\bigg]\nonumber.
\end{align}
Again, the Gamma function in the denominator prevents $m$ from going above $m>1$.  The other components of the spinning four-point can be obtained in an analogous way. Let us also emphasize that the exchanged operators in this four-point function are the current itself and the scalar operator. 
The last step to obtain $Q_m$ is the gluing of the R-symmetry structures which is implemented by (see \cite{Goncalves:2019znr,Goncalves:2023oyx} for a detailed discussion about these formulas)
\begin{align}
&Y_{i,12}\times Y_{r,kj}    \rightarrow y_{1k}y_{2j}-y_{1j}y_{2k}\;,
\end{align}
where $i$ and $r$ are the points being glued.

The four-point function involving the stress tensor has a simple structure. The independent components are
\begin{align}
&M^{22}_L=\frac{16y_{12} y_{13} y_{23}}{3}\bigg(\frac{1}{\delta _{13}+\delta _{23}-1}-\frac{1}{\delta _{13}-1}-\frac{1}{\delta _{23}-1}-1\bigg)\;,\\
&M^{23}_L=\frac{32y_{12} y_{13} y_{23}}{3} \bigg(\frac{2}{\delta _{13}-1}-\frac{1}{\delta _{23}-1}-\frac{2}{\delta _{13}+\delta _{23}-1}-1 \bigg)\;,\nonumber
\end{align}
where the other components can be obtained by symmetry. In the same way we can obtain $L_m^{ab}$ for the exchange of stress tensor
\begin{align}
    L_{m}^{22}=\frac{16y_{12}y_{13}y_{23}}{3m!\Gamma(2-m)}\bigg(\frac{m-1}{\delta _{12}-1}+\frac{m-1}{\delta _{13}-1}+\frac{m-1}{\delta _{23}-1}-1\bigg)\;,\\
L_{m}^{23}=\frac{32y_{12}y_{13}y_{23}}{3m!\Gamma(2-m)}\bigg(\frac{2 (1-m)}{\delta _{12}-1}+\frac{2 (1-m)}{\delta _{13}-1}+\frac{m-1}{\delta _{23}-1}-1\bigg)\;,\nonumber
\end{align}
which allows us to conclude that there is a truncation. Let us add that $\tilde{L}_{m}$, defined in (\ref{eq:Lmtil})  can be obtained from the expressions above and it is obvious that it truncates. Note that the four-point function with one stress tensor can only exchange the scalar operator. This OPE makes the $4-4$ stress tensor factorization also sensitive to the $3-5$ scalar factorization, which gives a check of the consistency of both factorizations. 

\section{Rewriting AdS amplitudes in position space}
In this paper we completely determined the six-point function of $20'$ operators in the supergravity approximation without computing explicitly any Witten diagram. Moreover, the algorithm is entirely within Mellin space. Nevertheless, it might still be useful to write the result in terms of position space functions. The goal of this appendix is to provide a map between different terms in our Mellin amplitude and functions in position space, and can be read independently from the rest of the paper.

By using the integrated vertex identities (see, e.g., Appendix C.2 of \cite{Bissi:2022mrs}), one can integrate out particle exchanges and reduce all Witten diagrams to two types. These are six-point contact diagrams (also known as $D$-functions) and 3-to-3 exchange Witten diagrams. As it will become clear in Mellin space, these two types of Witten diagrams further reduce to two seed functions which are denoted below as $D_{111111}$ and $I_{123,456}$.
\subsection{Six-point $D$-functions}
The $D$-function is defined as the following AdS integral
\begin{align}
D_{\Delta_1, \dots, \Delta_n} = \int \frac{dz_0 \, d^d z}{z_0^{d+1}} \prod_{i=1}^n \left( \frac{z_0}{z_0^2 + (\vec{z} - \vec{x}_i)^2} \right)^{\Delta_i}\;,
\end{align}
which has a constant Mellin amplitude. When dressed with factors of $x_{ij}^2$, we can obtain either polynomials or poles in the Mellin amplitude depending on the power of the factors. More precisely, this is achieved with the formula 
\begin{align}
    \prod_{i<j} (x_{ij}^2)^{-\alpha_{ij}} D_{\Delta_1 \dots\Delta_n} =
\int [d\delta] \prod_{i<j} (x_{ij}^2)^{-\delta_{ij}} \Gamma(\delta_{ij})
\left(
\frac{\pi^h \Gamma\left(\frac{\sum \Delta_i - d}{2}\right) \prod_{i} \Gamma(\delta_{ij} - \alpha_{ij})}{\prod_i \Gamma(\Delta_i) \Gamma(\delta_{ij})}
\right)\;,
\end{align}
which gives the Mellin amplitude
\begin{align}
    \mathcal{M}(\delta_{ij}) = 
\frac{\pi^{\frac{d}{2}} \Gamma\left(\frac{\sum \Delta_i - d}{2}\right)}{\prod_i \Gamma(\Delta_i)} 
\prod_{i<j} \frac{\Gamma(\delta_{ij} - \alpha_{ij})}{\Gamma(\delta_{ij})}.
\end{align}
It is then clear that these dressed $D$-functions can provide the singularities for all terms involving $P_{1,2,3,4}$ in (\ref{eq:ansatz6pt}). A useful property of these $D$-functions are the differential recursion relations
\begin{align}
    D_{\Delta_1,\dots,\Delta_i+1,...,\Delta_j+1,...,\Delta_n }= \frac{d-\sum_i \Delta_i}{2\Delta_i\Delta_j} \frac{\partial^2}{\partial x_{ij}^2} D_{\Delta_1,\dots,\Delta_n}\;,
\end{align}
which relate $D$-functions with different weights. In fact, in our case all the $D$-functions can be reduced by these relations to $D_{111111}$ only. This basic $D$-function can also be represented as a conformal one-loop integral in six dimensions \cite{Penedones:2010ue,Paulos:2012nu}
\begin{align}
D_{111111} = \int \frac{d^dx_0}{x_{10}^2x_{20}^2x_{30}^2x_{40}^2x_{50}^2x_{60}^2}\;. 
\end{align}
This integral has recently been computed and expressed in terms of classical polylogarithms with weight $3$ \cite{Ren:2023tuj} and obtained around $3$ non-consecutive lightcones in \cite{DelDuca:2011wh}\footnote{In particular is useful for the triple snowflake pole piece in the ansatz.}. In fact, there are some similarities between the integrand for $3$ lightcone limit of $D_{111111}$ and the integrand for the six-point lightcone conformal blocks in the snowflake channel. In \cite{DelDuca:2011wh} it was found that the following parametrization of the cross ratios
\begin{align}
    &u_2= -\frac{(x_8-1) (z_2-z_5)}{z_2 (x_5-x_8)}, \ \ u_4= -\frac{y_2-y_8}{(x_8-1) (y_2-1)}, \ \ u_6= \frac{y_8 z_2}{y_8-y_2},\\
    &U_1= -\frac{(x_8-1) (z_5-1)}{x_8-x_5}, \ \ U_2= -\frac{z_2}{y_2-1}, \ \ U_3= \frac{x_5 (y_2-y_8)}{(y_2-1) (x_5-x_8)}\;,\nonumber
\end{align}
was useful to obtain an expression in terms of polylogarithms. It would be interesting to study these new parametrization for the snowflake conformal blocks.

\subsection{Exchange diagram: Six-point double box}
The remaining Witten diagrams have poles separating the external points into two groups of 3 and 3. These poles involve the sum of three $\delta_{ij}$ and cannot be generated from $D$-functions. However, they can be expressed in terms of the following two-loop six-point integral in four dimensional flat-space
\begin{align}
I_{123,456}=\int \frac{d^4x_7d^4x_8}{x_{17}^2x_{27}^2x_{37}^2x_{78}^2x_{48}^2x_{58}^2x_{68}^2}\label{eq:sixpointdoublebox}\;.
\end{align}
To our knowledge this integral has not been fully computed  in general kinematics. Its symbol is known \cite{Spiering:2024sea} and the integral is expressed in terms of elliptic functions\footnote{However, for specific configurations the integral can be expressed in terms of Goncharov polylogarithms. One such case is when all points are on a common line \cite{Rodrigues:2024znq}. }.

The Mellin representation of this Feynman integral is given by \cite{Paulos:2012nu}
\begin{align}
    I_{123,456}=\int [d\delta] \frac{1}{\delta_{12}+\delta_{13}+\delta_{23}-1}\prod_{i<j}\frac{\Gamma(\delta_{ij})}{(x_{ij}^2)^{\delta_{ij}}}\;,
\end{align}
where the Mellin variables satisfy $\sum_{j}\delta_{ij}=0$ with $\delta_{ii}=-1$. Similar to the $D$-functions discussed in the previous subsection, any Mellin amplitude of the comb type present in our result can be expressed in terms of the derivatives of $I_{123,456}$ with respect to $x_{ij}^2$ and multiplication by their powers. To see this more explicitly, let us define 
\begin{align}
A(x_{ij}^2)=\int [d\delta_{ij}] M(\delta_{ij}) \prod_{1\leq i <j\leq n}\Gamma(\delta_{ij}) (x_{ij}^2)^{-\delta_{ij}}\;,\label{eq:MellinDefinition}
\end{align}
as a generic conformal function of $n$ points written in Mellin space. 
Then the Mellin amplitude of the derivative of (\ref{eq:MellinDefinition}) with respect to $x_{mn}^2$ is given by
\begin{align}
\frac{\partial }{\partial x_{mn}^2} A(x_{ij})  = \int [d\delta] M(\delta_{ij}'-\delta_{i}^{m}\delta_{j}^{n}-\delta_{i}^{n}\delta_{j}^{m}) \, \Gamma(\delta_{ij}') (x_{ij}^2)^{-\delta_{ij}'}\;,
\end{align}
where the Mellin variables satisfy $\sum_{j}\delta_{ij} = \Delta_i +\delta_{i}^{m}+\delta_{i}^{n}$. On the other hand, the Mellin amplitude obtained by multiplying a power of $x_{ij}^2$ is given by
\begin{align}
(x_{mn}^2)^{-\alpha_{mn}}A(x_{ij}^2) = \int [d\delta] M (\delta_{ij}'-\alpha_{mn}\delta_{i}^{m}\delta_{j}^{n}) (\delta_{ij}')_{-\alpha_{mn}}  \Gamma(\delta_{ij}')\;. 
\end{align} 
For example, using these formulas we can obtain the position space representations for the following Mellin amplitudes
\begin{align}
    &\frac{2x_{16}^2x_{35}^2}{x_{12}^2x_{34}^2}\partial_{x_{13}^2}\partial_{x_{56}^2}^2 I_{126,345}\rightarrow  -\frac{\delta _{16} \delta _{35}}{\left(\delta _{12}-1\right) \left(\delta _{34}-1\right) \left(\delta _{35}+\delta _{45}\right)}\;,
\\
    &\frac{x_{35}^2}{x_{12}^2x_{34}^2}\partial_{x_{35}^2}\partial_{x_{56}^2} I_{126,345}\rightarrow \frac{\delta _{35}}{2 \left(\delta _{12}-1\right) \left(\delta _{34}-1\right) \left(\delta _{35}+\delta _{45}-1\right)}\;.
\end{align}

\bibliography{refs} 
\bibliographystyle{utphys}

\end{document}